\documentclass[fleqn,10pt]{wlscirep}
\usepackage[utf8]{inputenc}
\usepackage[T1]{fontenc}
\usepackage{lineno}
\usepackage[center]{subfigure}

\usepackage{comment}
\usepackage{hyperref}

\title{Understanding the Vegetable Oil Debate and Its Implications for Sustainability through Social Media}

\author[1,2]{Elena Candellone}
\author[3,4]{Alberto Aleta}
\author[1,5]{Henrique Ferraz de Arruda}
\author[6,7]{Erik Meijaard}
\author[3,4,5]{Yamir Moreno}
\affil[1]{ISI Foundation, Torino, Italy}
\affil[2]{Department of Methodology and Statistics, Utrecht University, Utrecht, The Netherlands}
\affil[3]{Institute for Biocomputation and Physics of Complex Systems, University of Zaragoza, Zaragoza, Spain}
\affil[4]{Department of Theoretical Physics, Universidad de Zaragoza, Zaragoza, Spain}
\affil[5]{CENTAI Institute, Turin, Italy}
\affil[6]{Borneo Futures, Bandar Seri Begawan, Brunei}
\affil[7]{The University of Queensland, School of Biological Sciences, Brisbane, QLD, Australia}

\begin{abstract}
The global production and consumption of vegetable oils have sparked several discussions on sustainable development. This study analyzes over 20 million tweets related to vegetable oils to explore the key factors shaping public opinion. We found that coconut, olive, and palm oils dominate social media discourse despite their lower contribution to overall global vegetable production. The discussion about olive and palm oils remarkably correlates with Twitter's growth, while coconut increases more significantly with bursts of activity. Discussions around coconut and olive oils primarily focus on health, beauty, and food, while palm draws attention to pressing environmental concerns. Overall, virality is related to environmental issues and negative connotations. In the context of Sustainable Development Goals, this study highlights the multifaceted nature of the vegetable oil debate and its disconnection from scientific discussions. Our research sheds light on the power of social media in shaping public perception, providing insights into sustainable development strategies.
\end{abstract}

\begin{document}

\flushbottom
\maketitle

\thispagestyle{empty}

\section*{Introduction}

In 1987, the World Commission on Environment and Development, commonly known as the Brundtland Commission, published the report \emph{Our Common Future}. This report included the definition of \emph{Sustainable development} as the one ``that meets the needs of the present generation without compromising the ability of future generations to meet their needs'', and it was further subdivided into three pillars or dimensions: environment, society, and economy \cite{Secretary-General1987Aug,Purvis2019May}. In the following decades, sustainable development grew in relevance as one of the main organizing principles for meeting human development goals, leading to the adoption of the 2030 Agenda for Sustainable Development by all United Nations Member States in 2015\cite{UN_SGD}.

Vegetable oils are currently one of the major elements in the global food system. Their global demand has increased dramatically since 1960, with a 10-fold increase in soybean and a 25-fold increase in palm oil consumption by 2010 \cite{Byerlee2016Nov_1}. Vegetable oils are an important source of calories, especially in the developing world. They can be found in more than 50\% of packaged products in supermarkets and are also good for animal feed and biofuel production\cite{Yue2020May}. Besides, they have generated huge economic benefits for their producers \cite{Byerlee2016Nov_2}. For instance, in 2016, soybean constituted about one-third of the agricultural export earnings of Brazil, and oil palm accounted for half of Indonesia's agricultural export earnings. 

Growing oil demand has primarily been met by increasing oil production areas. However, most crop expansion has taken place in very few countries, especially around the tropics. In 2019, Indonesia and Malaysia accounted for 84.4\% of the world's palm oil production, while Argentina and Brazil accounted for 60\% of global soybean oil exports \cite{CME_group,Phalan2013Jan}. The expansion of cultivated land in these countries is associated with dire environmental impacts, such as deforestation and threats to biodiversity \cite{Carlson2018Jan}. Yet, even though palm oil accounts for 40\% of global vegetable oil demand, it covers less than 5.5\% of the total global oil crop area thanks to its large yield per hectare \cite{Meijaard2020Dec}. Thus, meeting the same demand with other options may be even worse in terms of environmental impact \cite{Parsons2020Jun}. As a consequence, there is an ample debate among the scientific community on the role of vegetable oils in the context of the Sustainable Development Goals (SDGs) \cite{Meijaard2020Dec,obaideen2020,Santika2019Aug,Alcock2022Jul}. In this paper, our aim is to understand how this debate is perceived by the public on social media. 

Social platforms have played an important role in shaping public opinion on vegetable oils since their early beginnings. In March 2010, the environmental protection group Greenpeace kicked off a social media campaign against one of the largest food processing companies in the world, Nestlé. The campaign revolved around the use of palm oil in one of their best-known products, claiming that its production led to the destruction of rainforests. In spirit, it was similar to the one held in 2008 against Unilever, which was the target of 'raids' by Greenpeace activists dressed as orangutans (Pongo spp.) in several facilities. This time, however, the campaign was initially fully virtual, starting with a video on Youtube and then moving to Nestlé's Facebook page when the video was taken down. In just a few days, the campaign jumped to Twitter and the mainstream news headlines. The reputation crisis led Nestlé to suspend purchases from one of their suppliers, join several sustainability initiatives, and promise an analysis of their whole supply chain. This marked a new era for social protests on corporate practices \cite{Debapratim,Champoux2012Mar}. 

Given this context, we focus our analysis on the microblogging platform Twitter, which is commonly used by companies to build reputation, disseminate word-of-mouth campaigns and shape the company's image \cite{Killian2015Sep, Srivastava2020Oct}. On this platform, users may follow other members without the need to reciprocate. This asymmetry facilitates the creation of opinion relationships and communities \cite{Gruzd2011Jul}. Besides, its wide audience from various backgrounds and geographical locations is especially interesting in studying global opinion trends \cite{DeLima2022Oct}. This characteristic of the platform is unfortunately leveraged by certain groups to spread false news and misinformation \cite{Vosoughi2018Mar,Varol2017May}. 

In spite of the important relationship between public debate and social networks, the literature on vegetable oils and social media is relatively small. Pradipta and Jayadi \cite{pradipta2022} tested two sentiment analysis techniques with a small dataset of 26,469 tweets on palm oil collected between July and September 2021, but did not interpret their output. The media analytics company Commetric published a report on the brands most often referenced in the palm oil debate based on 4,503 tweets collected between September 2018 and September 2019 \cite{commetric}. They found a low overall engagement with palm oil news, with bursts of activity caused by stories about the damage of palm oil production to orangutan habitats. Ruggeri and Samoggia \cite{Ruggeri2018Jan} analyzed 16,764 tweets collected from 198 palm oil agri-food chain companies. They observed that palm oil producers actively used Twitter to promote palm oil sustainability, while European manufacturers and retailers limited their activity to react to consumers' questioning. Khairiza and Kusumasari discussed the effect of a palm oil social media campaign from the Indonesian government using 9,628 tweets collected from September to November 2019 \cite{Khairiza2020Sep}. Finally, van Rossum \cite{rossum2018} conducted an analysis of Twitter within the framework of Habermas's concept of the public sphere. Utilizing a case study of 103,500 tweets about palm oil collected from July to September 2018, the study concluded that Twitter does not provide the ideal platform for the formation of public opinion as envisaged in Habermasian terms.

Outside of Twitter, Teng et al. \cite{Teng2020Nov} explored 4,260 posts from Youtube and Reddit on palm oil. Other authors examine the value social media can provide as a marketing tool \cite{Agustin2021Dec} or for agricultural technology and information expansion \cite{AbRahman2021Dec,Peng2021} in the context of palm oil. Notably, all these references revolve around palm oil, neglecting the contribution of other vegetable oils to the broader debate on the sustainability of the global food chain. 
Interestingly, in an investigation by Alvarez-Mon et al. on social attitudes towards the Mediterranean diet, based on 1,608 tweets published by 25 US media outlets between January 2009 and December 2019, olive oil was the element of the diet that generated the least attention \cite{Alvarez-Mon2022Jan}. To conclude this brief overview, it is worth mentioning that in De Lima \cite{DeLima2022Oct} and references therein, they analyze the discussion on topics related to sustainability in Twitter, but none of them explicitly mention vegetable oils. 

\section*{Results}
\subsection*{Vegetable oils presence in Twitter}

\begin{figure}
    \centering
    \includegraphics[width=\textwidth]{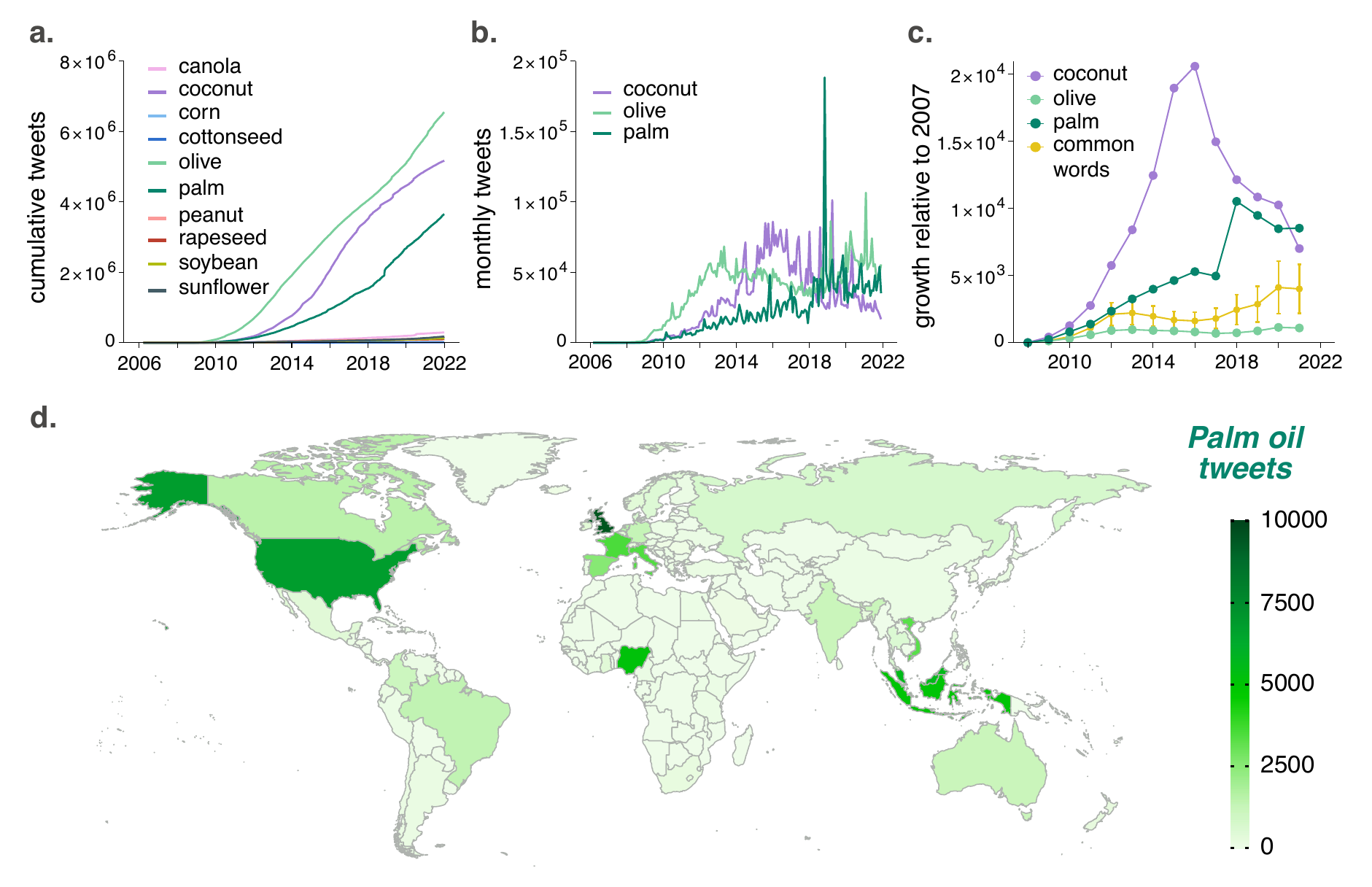}
    \caption{\textbf{Anatomy of the vegetable oils presence in Twitter:} tweets sent from 3/21/2006 to 12/31/2021 containing the bigram ``\emph{type oil}'', where \emph{type} corresponds to any of the oils listed in the legends. (a) The cumulative number of tweets for each oil. There are three major oils in terms of social media presence: olive, coconut, and palm oils. (b) The monthly number of tweets for each of the three major vegetable oils. (c) Growth relative to 2007, measured as the total number of tweets in a given year over the number of tweets in 2007. We compare the growth of tweets on each vegetable oil with the growth of the 100 most common words in English as a proxy for the growth of Twitter (see Methods). Error bars indicate the standard deviation of the growth for the common words. (d) The number of geo-tagged tweets per country containing the bigram ``palm oil'' in 102 languages (see Methods). The total number of tweets collected in any language is 7,946,915, of which only 64,682 are geo-tagged. }
    \label{fig:anatomy}
\end{figure}

We begin the characterization of the vegetable oils debate on Twitter by looking at the cumulative number of tweets associated with 10 major oils, Fig.~\ref{fig:anatomy}a (see Methods). We observe two clear groups of oils: coconut, olive, and palm oil, with 5 million, 6.5 million, and 3.7 million tweets, respectively, and the rest with fewer than 300,000 tweets each (see Supplementary Information for more details). Note that this is not proportional to the production volumes of these oils. The world supply of olive and coconut oils is lower than 4 million metric tons. In contrast, the world supply of palm, soybean, and sunflower oils in 2022 was 76, 59, and 20 million metric tons respectively, \cite{oilSeeds} with soybean, with 130 million hectares of land, having by far the largest environmental footprint of all oil crops \cite{faostat2023}. Thus, the major oils in terms of social media presence are coconut, olive, and palm, regardless of their actual supply. Henceforth, we focus our analysis on these three oils.

In Fig.~\ref{fig:anatomy}b, we show the monthly number of tweets of the three major oils. Olive oil is characterized by a quick upwards trend until 2013, followed by a slow decay which reverts in 2018. Furthermore, there are few bursts of activity, which are indicators of viral events. Coconut oil has a smaller initial growth, but between 2014 and 2018 is the oil with the most monthly tweets. The interest in coconut oil decayed from the peak of 2016, with some noticeable bursts until 2022. Lastly, palm oil is characterized by a considerably slower but constant upwards trend, with a major viral event in November 2018 with twice as many tweets as the largest burst of any of the other two oils analyzed.

The increments in the number of tweets could mean either that these topics capture more public attention or they might be a direct consequence of the own growth of the platform. To elucidate this, in Fig.~\ref{fig:anatomy}c, we show the relative growth in the number of tweets for each oil since 2007, compared to the growth of tweets containing any of the 100 most common words in English as a proxy for Twitter's growth (see Methods). At first glance, the overall trend of tweets containing common words is similar to the palm and olive oils tweets.
Through the Granger causality test, we confirm that the temporal trend of common tweets significantly predicts the trends in tweets about palm and olive oils (p-value $<10^{-10}$). This predictive association, however, does not extend to coconut oil tweets (p-value $0.09$). These findings suggest that a notable proportion of the rise in tweet volumes can be attributed to the general expansion of Twitter's user base, rather than specific increases in interest or discussions about these oils. Further details are available in the Supplementary Information.

The growth of olive oil tweets is smaller than the average growth of common tweets, signaling that it is a topic that does not gather special attention. Besides, the decay from 2013 to 2018, followed by an upward trend, is similar to the one seen in common words, reinforcing the idea that its evolution does not differ from the average conversation on the platform. Conversely, the conversations on coconut and palm oils grow much more than common tweets, indicating an increase in public interest. However, coconut oil shows a very fast growth only until 2016, the point at which it starts to decline steadily.

To conclude this initial characterization, we collected all tweets containing the bigram ``palm oil'' written in 102 languages (see the Supplementary Information for a complete list of the bigrams). There are 7,946,915 tweets in this set, of which roughly 50\% are in English. In Fig.~\ref{fig:anatomy}d, we show the number of tweets that are geo-tagged in each country (see Methods). As expected, given the origin of the aforementioned campaign, the country with the largest number of tweets is the United Kingdom, followed by the United States, Malaysia, Nigeria, and Indonesia. Notably, Nigeria used to be the largest producer of palm oil in the world in the 1960s, and it is currently the largest consumer in Africa \cite{PricewaterhouseCoopers2022Nov}. Thus, the debate is mainly centered in the UK and the US, together with countries that are historical palm oil producers and major consumers.

\subsection*{The debate around vegetable oils}

\begin{figure}
    \centering
    \includegraphics[width=\textwidth]{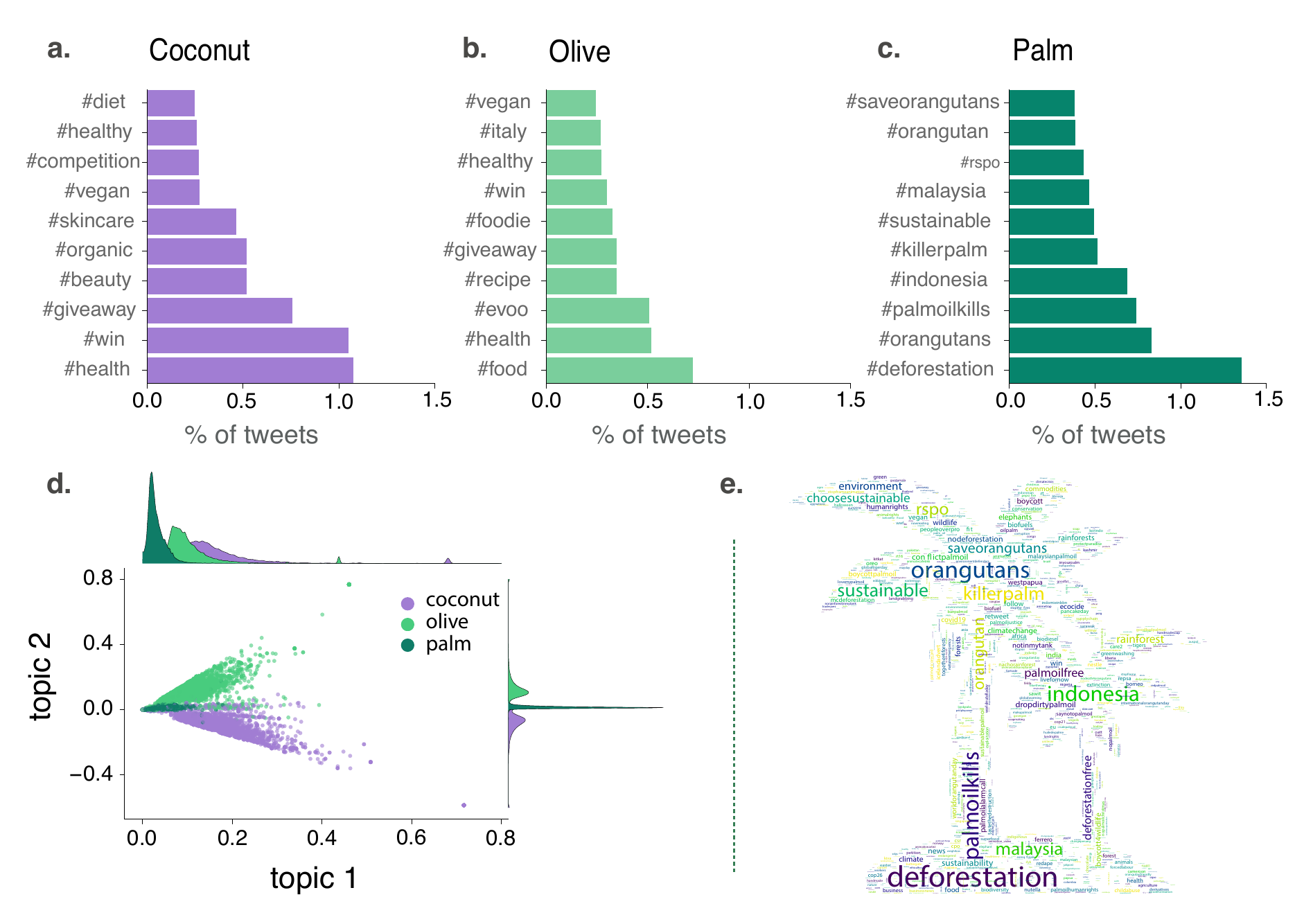}
    \caption{\textbf{The debate around vegetable oils:} the topics associated with each vegetable oil can vary greatly. Panels a-c show the top 10 most used hashtags for each set of tweets, expressed as the percentage of tweets containing the hashtag, for coconut, olive, and palm oil, respectively. Palm oil is mainly related to environmental issues, while coconut and olive are predominantly associated with nutrition and health topics. Panel d reports the results of a Latent Semantic Analysis (LSA) applied to the set of tweets (see Methods). Each point in the plot represents a tweet, with color indicating the vegetable oil it is associated with. The closer the two points are in space, the more similar their topics are. Lastly, panel e depicts a word cloud of the top 2,000 hashtags in the palm oil dataset. Font size is proportional to the number of occurrences in the dataset.}
    \label{fig:debate}
\end{figure}

Next, we look at the content of tweets to characterize the debate. First, we extract the top 10 hashtags in each set of tweets in English (see Methods). These hashtags can be used as a proxy for the most common topics associated with each oil. In Fig.~\ref{fig:debate}, we show the percentage of tweets that contain each of these hashtags. For coconut oil, we find that most hashtags are related to health or beauty. There is also an important presence of tweets containing \emph{giveaway} or \emph{win}, which could be related to marketing campaigns. The topics around olive oil are similar, although there is a larger presence of keywords related to food and nutrition instead of beauty. In contrast, in the case of palm oil, most tweets are related to the negative environmental effects associated with it. Most of them are against palm oil production, except for \emph{sustainable} and \emph{rspo}, which stands for Roundtable on Sustainable Palm Oil, an organization with the objective of promoting the growth and use of sustainable palm oil.

Not all tweets contain hashtags. We only find hashtags in 17.7\%, 19.5\%, and 25.9\% of the tweets in the sets of coconut, olive, and palm oil, respectively. To broaden the analysis, we apply Natural Language Processing (NLP) techniques to visualize tweets according to their main topics using the whole text rather than just hashtags. In particular, we apply a Latent Semantic Analysis (LSA), which projects the corpus of tweets into a 2-dimensional space. In this projected space, if two points are close, it implies that their topics are closely related (see Methods). As we can see in Fig.~\ref{fig:debate}d, coconut and olive oil tweets have a wide distribution of topics. In contrast, the diversity of topics in tweets related to palm oil is much smaller, with most of them concentrated around the same axis. This confirms the results of the hashtag analysis. While coconut and olive oil are associated with beauty/food and health, the palm oil debate on Twitter is mostly focused on negative environmental impacts of its production.  This is further validated by Fig.~\ref{fig:debate}e, which shows a word cloud of the top 2,000 hashtags for the palm oil dataset. Most hashtags, even beyond the most common ones, are related to environmental issues.

We extend this analysis by looking at the sentiment associated with those topics. We employ a model trained with Twitter data for sentiment analysis. This technique estimates whether the opinion expressed in a text is positive, negative, or neutral (see Methods). In Fig.~\ref{fig:sentiment}, we report the fraction of tweets of each oil associated with positive, negative, or neutral sentiments. The fraction of neutral tweets is close to 50\% in the three cases, a value that is commonly found in sentiment analysis of Twitter independently of the specific topic \cite{Becken2019May,Gohil2018Apr,Becken2017Dec}. In contrast, the fraction of tweets associated with negative sentiments in the palm oil dataset is 4 times larger (42\%) than in the coconut (12\%) or olive oils (10\%). If we look at the evolution of tweets in each of these categories (panels d-f), we observe that their distributions are relatively stable across time (see also Fig.~S4 in the Supplementary Information) except for some viral events.

In the context of sustainability debates, we concentrate on the case of palm oil in 2018 due to two significant events that occurred during this year: the release of a report on palm oil's impact on biodiversity by the International Union for Conservation of Nature (IUCN) \cite{iucn2018}, and a widespread social media campaign by Iceland Foods and Greenpeace (see Discussion). Fig.~\ref{fig:palm_debate} presents a detailed breakdown of these dynamics. Panel \ref{fig:palm_debate}a demonstrates that the IUCN report, despite its scientific significance, did not trigger a substantial public response, with tweets mentioning the IUCN peaking at a modest 2,500 in the week of the report's release before rapidly fading away. Intriguingly, the usage of the term 'biodiversity' shows a high correlation with the appearance of 'IUCN', suggesting that public discussions around biodiversity were mainly restricted to the context of the IUCN report. This indicates a disconnection between the scientific and public discourses on this matter, as the issue of biodiversity seemed largely absent from the broader public debate.

In stark contrast, the social media campaign involving Iceland Foods and the symbol of the orangutan spurred a much larger response, as reflected in the volume of related tweets which reached nearly 25,000 during the campaign's peak. This illustrates the powerful role of emotive campaigns in sparking public interest and engagement. Panels \ref{fig:palm_debate}b and \ref{fig:palm_debate}c provide an insight into the emotional tone of these debates. Most tweets related to the IUCN report were neutral, likely reflecting the typically unbiased language used by scientific agencies when disseminating information. However, a considerable portion was also negative, possibly signifying a public pushback against the report's findings, while positive tweets were conspicuously sparse. In contrast, the sentiment analysis for the social media campaign revealed a different pattern. The majority of tweets were negative, indicating a strong public reaction against the use of palm oil due to its ecological impacts. Positive tweets were also present, while neutral ones were very few, highlighting the strong emotional response triggered by the campaign. This suggests that emotive content, can play a critical role in shaping public opinion and engagement on sustainability issues.

\subsection*{Emergence of viral events}

Information often spreads through social networks as an avalanche. Users may see some information and desire to share it or comment on it. Then, an avalanche spreads if other users continue to spread that content \cite{Gleeson2016May}. Some topics may be commonly discussed, reaching a large volume of tweets over a long period. Conversely, certain ideas may spread very fast and reach comparable volumes but in a much shorter time. These are known as viral events - events that propagate widely and rapidly \cite{oxford_viral}. It is thus essential to study simultaneously both dimensions of the process: (i) the speed of the diffusion and (ii) its reach. To do so, we extract the set of tweets containing each of the 10 most common hashtags concerning each oil. Then, we measure the interevent time (IET), defined as the time interval between two consecutive tweets containing the same hashtag. To study the reach of a hashtag, we look at the daily number of tweets associated with it. We call this quantity the cascade size (CS) (see Methods for further details).

Since we have over 15 years of data, a certain hashtag may participate in multiple viral events. Hence, rather than focusing on a particular period, we study the overall IET and CS distributions for each hashtag under consideration. We fit these distributions using power-law functions ($p(x) \sim x^{-\alpha}$)characterized by the scaling parameter, $\alpha$, which are commonly found in any human activities \cite{Clauset_2009} (see Methods). In Fig.~\ref{fig:viral}, we represent the scaling parameter of the CS distribution, $\alpha_\text{CS}$, versus the IET distribution, $\alpha_\text{IET}$. These functions exhibit, in the limit $x\rightarrow\infty$, a well-defined mean only when $\alpha > 2$, and a well-defined variance when $\alpha > 3$. For finite systems, when $\alpha < 3$, the average fails to capture the dynamics accurately due to pronounced fluctuations. These distinct regimes form the foundation for our definition of a virality phase diagram as a function of $\alpha$. Hashtags can then be characterized according to the area in which they lay in this plane.

As we can see in Fig.~\ref{fig:viral}, the hashtags associated with each type of oil tend to fall into similar categories. For instance, olive oil tweets are mostly found in regions III and VI, characterized by well-defined volumes and relatively slow dynamics. In contrast, most palm oil tweets are in Region II, which we call the viral region. Hashtags in this region have well-defined CS and IET averages, but their respective variances diverge. This implies that these events do not have easily predictable dynamics, as it is possible to have some cascades with just a few tweets, followed by a period of silence, and then a large burst of tweets in a short period. Interestingly, we observe that the hashtags associated with this region are the ones most related to environmental issues, especially those with negative connotations. The few hashtags in Region V, with well-defined sizes but unpredictable interevent times, are those supporting palm oil, such as ``rspo'' or ``sustainable''. 

\begin{figure}
    \centering
    \includegraphics[width=\textwidth]{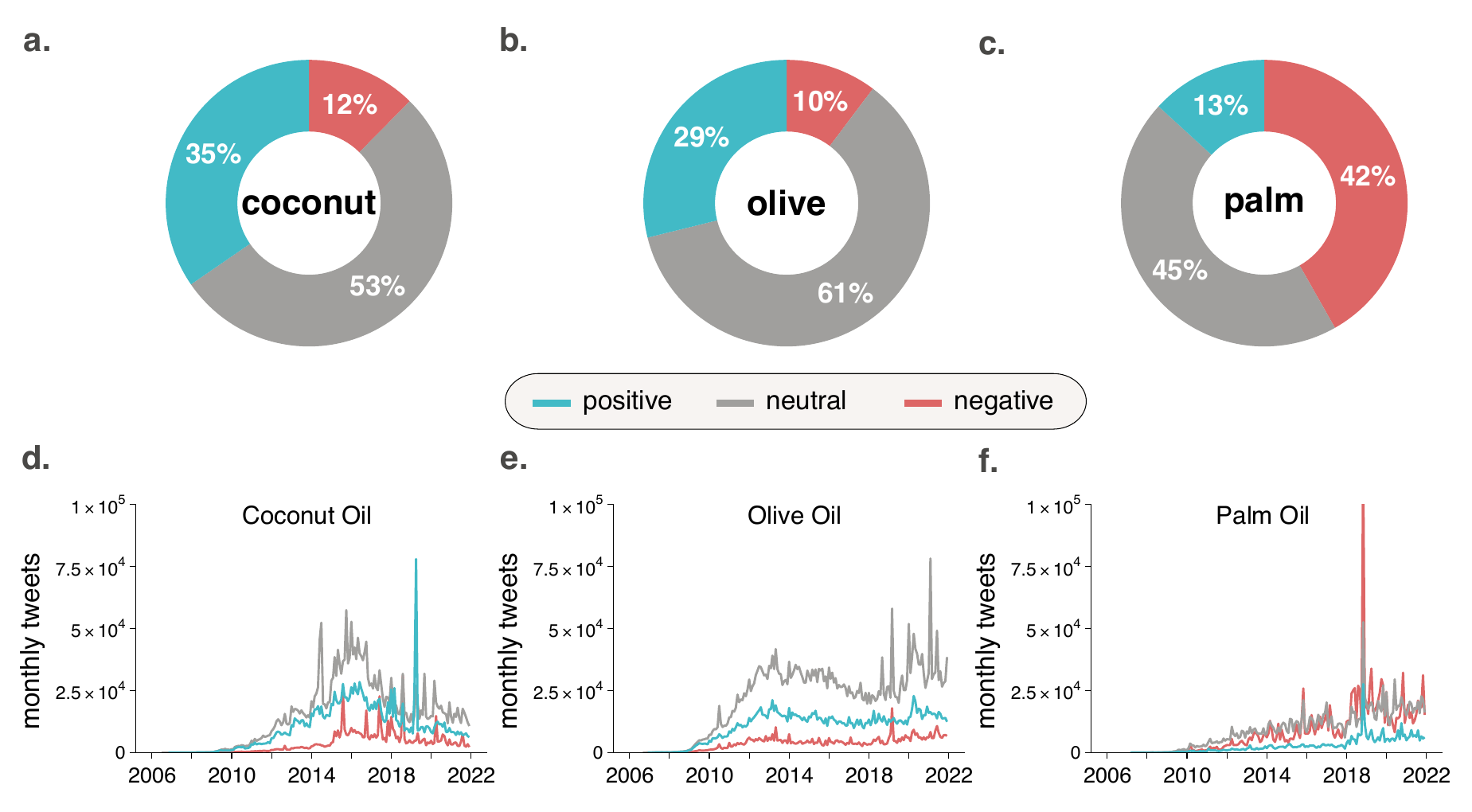}
    \caption{\textbf{Sentiment Analysis:} Panels a-c contain the fraction of tweets associated with each sentiment for each of the three datasets under consideration (coconut, olive, and palm). In the three oil cases, roughly 50\% of the tweets are considered neutral, but the situation is completely different for negative ones. The datasets of coconut and olive oils contain about 10\% of tweets labeled as negative, while this number increases to 42\% for palm oil. Panels d (coconut), e (olive), and f (palm) show the monthly number of tweets associated with each sentiment. The share of tweets in each category is fairly constant, except for a few viral events, such as the one related to coconut oil in March 2019 and the one associated with palm oil in November 2018.}
    \label{fig:sentiment}
\end{figure}

\begin{figure}
    \centering
    \includegraphics[width=\textwidth]{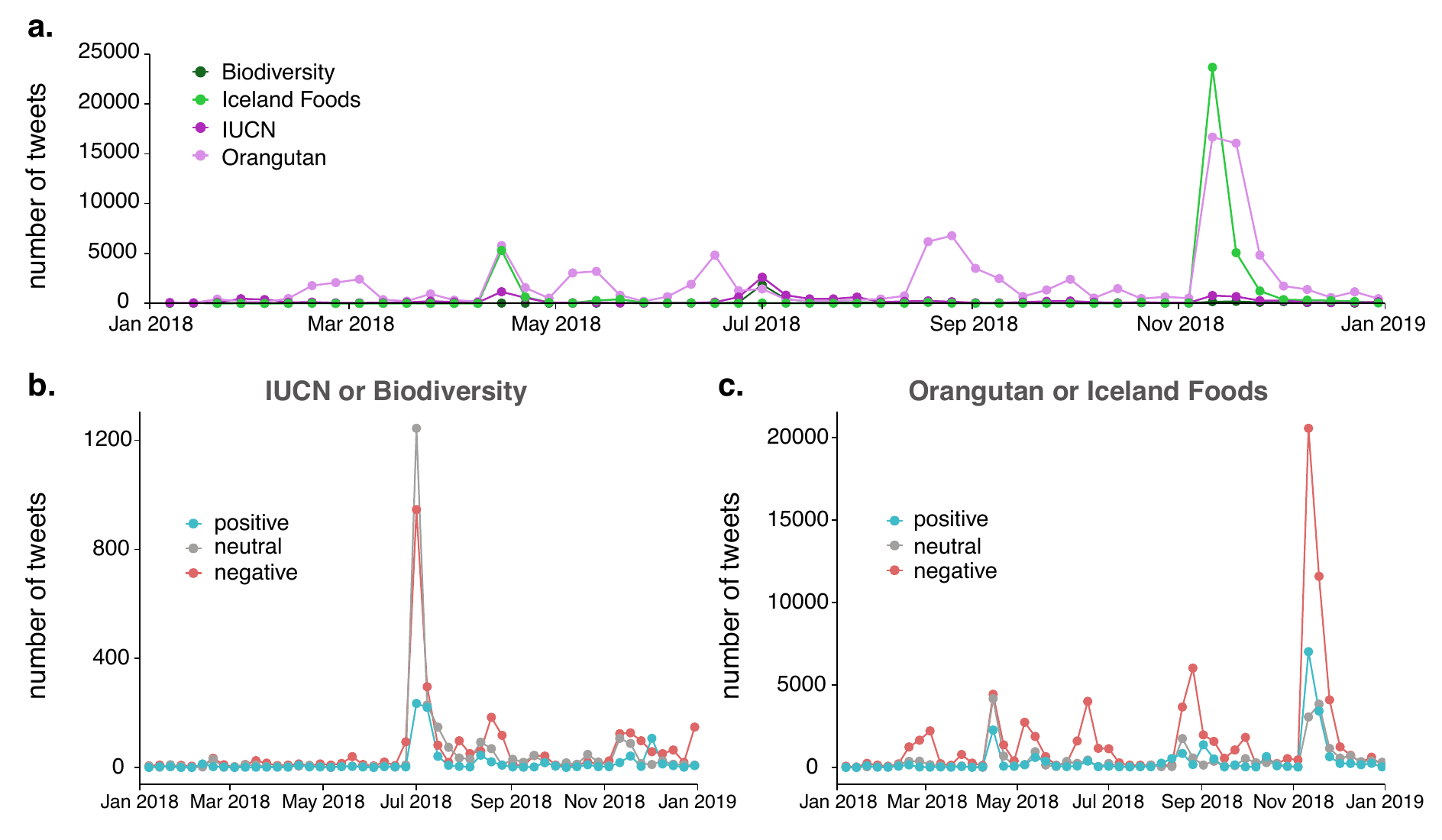}
    \caption{\textbf{The palm oil debate in 2018:} Panel (a) displays the weekly count of tweets containing the keywords 'Biodiversity', 'Iceland Foods', 'IUCN', and 'Orangutan', illustrating the varying public attention to different facets of the issue. This showcases two significant but differently echoed events: the IUCN report on palm oil's impact on biodiversity in late June and the widely noticed campaign against palm oil by Iceland Foods and Greenpeace in Christmas. Panel (b) visualizes the sentiment (positive, neutral, negative) associated with tweets containing the terms 'IUCN' or 'biodiversity', demonstrating the emotional response to scientific discussions. Panel (c) presents the sentiment associated with tweets containing the terms 'Orangutan' or 'Iceland Foods', reflecting the emotional impact of the viral campaign. }
    \label{fig:palm_debate}
\end{figure}

\begin{figure}
    \centering
    \includegraphics[width=\textwidth]{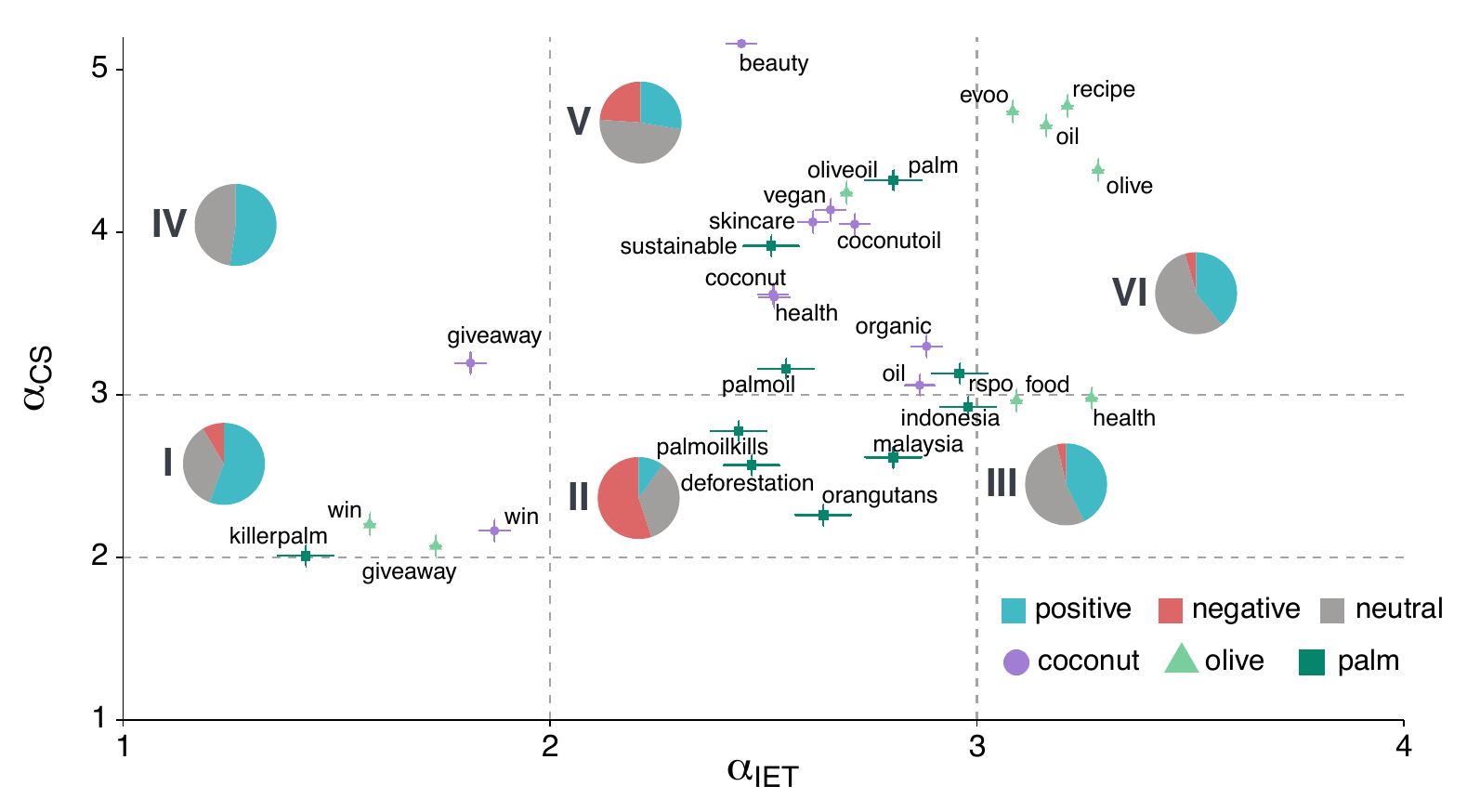}
    \caption{\textbf{Virality phase diagram:} Each point represents one of the 10 most common hashtags in the coconut (orange circles), olive (blue triangles) and palm (green squares) datasets. The spatial coordinates are the scaling exponents of the interevent time ($x$-axis) and the cascade size ($y$-axis) distributions. Gray lines delimit the areas defined by the critical exponents (see Methods). Pie charts show the distribution of positive, neutral and negative tweets with at least one hashtag in each of these areas. }
    \label{fig:viral}
\end{figure}

These observations are further supported by the sentiment distribution of the tweets associated with each region. Regions I and IV contain mostly positive tweets. As we can see, those hashtags, except for ``killerpalm'', are related to freebies given during marketing campaigns, which explains that sentiment. Besides, the short duration of these campaigns has an important effect on the IET distribution. The hashtag ``killerpalm'' shares these characteristics since it was related to a short-lived environmental campaign that took place during the 2015 United Nations Climate Change Conference (COP21) \cite{irishmedia}. The small fraction of negative tweets observed in this area belongs to this hashtag. In Region II, as previously discussed, we find most hashtags related to environmental issues and palm oil, which yields a large number of tweets with negative sentiment. In Region V, positive tweets tend to be associated with coconut oil, while a relatively large fraction of negative tweets are associated with palm oil. Lastly, in regions III and VI, since all of them are related to olive oil, most tweets are neutral, in agreement with Fig.~\ref{fig:sentiment}b.

\section*{Discussion}
The role of vegetable oils in the context of the Sustainable Development Goals is far from trivial, as they may have a positive impact on objectives such as no poverty (SDG 1) or zero hunger (SDG 2) but a negative one on climate action (SDG 13) or life on land (SDG 15). This complex relationship has sparked an important debate within the scientific community and, thanks to social media, also within the general public. However, it is crucial to comprehend the factors that influence public opinion and the role of social media in shaping public discourse. Only  then we would be in the position to use collected data for policymaking. 

Guided by the previous gap of knowledge, in this paper we have investigated a corpus of over 20,000,000 tweets on vegetable oils, including a subset of tweets in 102 languages published between March 2006 and December 2021. Our results indicate that discussions about coconut, olive, and palm oils are the most prevalent on Twitter, even though their actual supply is much lower than other oils, such as soybean or sunflower. In terms of growth, we observe that the evolution of olive oil tweets and palm oil tweets (at least until 2018) are highly correlated with the overall growth of Twitter, in contrast with coconut oil. In the early 2010s, coconut oil gained a lot of attention. Many labeled it as a ``superfood'' as many media outlets affirmed that it could help in weight loss, together with several health benefits \cite{Clegg2017Oct,Lockyer2016Mar}, which could explain the rapid growth in the number of tweets. But the fad started to decay after reaching its peak in 2016. Besides, the public perception of this oil may have changed after the publication of a review on dietary fats by the American Heart Association in 2017 \cite{Sacks2017Jul,Dewey2018Mar}, leading to its decline in terms of attention.

Similarly, palm oil shows stable growth, with few bursts of activity for over 10 years. The situation completely changed on November 9, 2018 \cite{Berg2022Nov}. Previously, in April, the UK supermarket company, Iceland Foods, announced that they would remove palm oil from all their products in an effort to fight against deforestation. The campaign reception was generally positive but relatively small. During summer, Iceland became aware of an animated campaign film made for Greenpeace featuring a baby orangutan that had lost its mother and its forest due to palm oil. They asked to use the film as their Christmas TV advertisement and booked airtime for November. However, one week before its release, the cable company banned the advertisement, claiming it was political. On November 9, when the advertisement should have been broadcast, the company released the commercial on its website and several social channels. They also explained that it had been banned from TV. This led to an extreme public reaction, with the video becoming viral, showcasing the Streisand effect (when an effort to censor information backfires by increasing awareness of that information) \cite{Jansen2015Feb}. This permanently altered the dynamics of the public debate, with the annual growth after 2018 well above the ones before the campaign. Notably, these discussions are mainly centered in the UK and the US, together with countries that are historical palm oil producers.

An analysis of the content of the tweets suggests that discussions about coconut and olive oils tend to focus on health and beauty for the former and on health and food for the latter. Conversely, discussions about palm oil are primarily centered on environmental issues. The sentiments associated with the first two oils tend to be positive, while tweets related to palm oil are mostly negative. These distributions are fairly constant across time, except for some viral events. For instance, in the case of coconut oil, there was a large spike of positive tweets in March 2019. This is due to a joke tweet that went viral, with over 86,000 retweets, which explains why the event is mostly positive. For palm oil, instead, there was a spike in all sentiments in November 2018. These tweets are related to the environmental campaign carried out by Iceland Foods and Greenpeace. Even though the overwhelming majority are labeled as negative, there are also some tweets in the other categories, signaling that the event spread through several conversations and points of view. Furthermore, the viral event of coconut oil did not have long-lasting impacts, while the one of palm oil altered public perception up to the last day of data (see Fig.\ref{fig:anatomy}c). The lack of social media impact from a key publication by the IUCN - the largest conservation organization in the world - is food for thought for scientists seeking to use science and social media to influence public perceptions on controversial and polarized topics. The powerful amplification potential of social media may not apply to nuanced scientific information possibly because herd behaviour on social media \cite{herd2020} is driven by emotive reactions to positive or negative rather than neutral messages. It is difficult to get emotional about a finding that oil palm can be good or bad depending on the context.

We conclude the analysis by focusing on the diverse characteristics of viral events concerning each oil. Topic popularity and interevent time distributions tend to be very fat-tailed \cite{Kivela2015Nov, Gleeson2016May, Brien2020Jun}, and this pattern is also present in the debate on vegetable oils. We observe events reaching a large number of people in a short period and others having a slower, more sustained spread. In particular, hashtags associated with olive oil generally have a slower spread and more predictable dynamics. In contrast, those associated with palm oil are more likely to be viral, with a larger variance in both the size and timing of their spread. We also verified a strong relationship between sentiment and virality. Hashtags related to environmental issues and negative connotations were more likely to be viral, and hashtags associated with marketing campaigns were more likely to have a faster spread but shorter duration. This has important implications since it has been observed that political messages containing moral-emotional language tend to spread more within ideological groups and less between them \cite{Brady2017Jul}.

In summary, these findings suggest that the debate about vegetable oils and SDGs on Twitter is multifaceted, with different oils being associated with different topics and sentiments. Yet, the interest in each oil and the specific topics involved are fairly disconnected from the scientific debate. Most oils go unnoticed in the overall discussion, while others are only associated with health or nutritional aspects. Only palm oil is discussed in terms of sustainability, and the conversations tend to be always negative, even though the cultivation of other oils may also have important environmental implications \cite{Meijaard2020Jul}. Indeed, it is crucial to note that the intensification of olive groves has posed a threat to biodiversity in Spain, resulting in the destruction of wintering bird communities' refuges \cite{Perez2023}. Furthermore, research has indicated that coconut oil is more dangerous for a higher number of species compared to other oils, followed by olive and palm oils \cite{Meijaard2020-ta}. In addition to the lack of topic variety, palm oil comes across as significantly more susceptible to viral events than other vegetable oils. This is another symptom of the lack of uniformity in the discussion around seemingly similar topics. Furthermore, the discussion around the sustainability of palm oil is an interesting example of how an awareness-raising campaign can bring the general public's attention to hitherto little-addressed issues and how the environmental issue is becoming preponderant in public discussion. As such, future awareness campaigns may want to highlight some of the strengths and weaknesses of this and other oils to provide consumers with the adequate tools to address issues as complex as the SDGs.  
    
\section*{Methods}

\subsection*{Basics of Twitter}

Twitter is a microblogging and social network platform in which users can post and interact with short messages known as tweets. Tweets were originally limited to 140 characters, but in November 2017, the limit was doubled to 280 characters. Pieces of text that start with \# are called hashtags, and usernames preceded by @ are known as mentions. These allow users to group their messages into topics and to address other users directly. Users may also follow the activity of other users, even though this relationship does not have to be reciprocal. Lastly, users may repost messages from other users and share them with their followers. This type of message is known as retweets \cite{ContributorstoWikimediaprojects2022Nov}.

\subsection*{Data collection}

We used Tweepy v4.4 \cite{roesslein2020tweepy} to connect to the Twitter API v2 with Academic Research access. We downloaded all tweets related to a selection of vegetable oils published between March 21, 2006 (the beginning of Twitter) and December 31, 2021. The Academic license allows access to the whole public tweets database, with a limit of 10 million tweets per month. We used as query bigram ``type oil'', where $type$ is either canola, coconut, corn, cottonseed, olive, palm, peanut, rapeseed, soybean, or sunflower. Note that queries are case-insensitive. This yielded a corpus of over 15 million tweets, which was later complemented with tweets containing the term ``palm oil'' expressed in 102 languages (see table S3 in the Supplementary Information for the exact terms), expanding the corpus to over 20 million tweets. Furthermore, according to the Oxford English Corpus, we also obtained the number of tweets per day containing any of the 100 most common words to estimate Twitter's growth (see table S4 in the Supplementary Information for the list of words).

\subsection*{Twitter analysis}
\textbf{Twitter growth:} for each oil, we compute yearly growth as the annual number of tweets over the same value in a reference year. In 2006, several oils were not discussed in the platform. As such, we take 2007 as the reference year. In the Supplementary Information, we show the results for other choices, yielding the same qualitative behavior (Fig.~S3).

\noindent
\textbf{Geolocation:} Twitter allows users to include their location in their tweets, which used to be a place or exact coordinates. Since 2019, the latter is no longer available. Besides, this option is disabled by default. As a result, it is estimated that less than 1\% of tweets are geo-tagged \cite{Kruspe2021Nov}. In our case, we found 64,682 geo-tagged tweets in the set of 7,946,915 tweets containing ``palm oil'' in any of the languages under consideration. We extracted the country of the place associated with the tweet directly from its metadata to estimate where they were produced.

\noindent
\textbf{Hashtags:} hashtags can be used to identify the most common keywords or topics related to a set of tweets. To obtain the most relevant hashtags, first, we remove the trivial hashtags in each set: \#oliveoil, \#olive and \#oil; \#coconutoil, \#coconut, and \#oil; and \#palmoil, \#palm, and \#oil. Then, we extract the 10 most common hashtags for each set of tweets. 

\subsection*{Latent Semantic Analysis}
Latent semantic analysis (LSA) is a natural language processing (NLP) technique based on the assumption that words that are close in meaning will occur in similar pieces of text. This allows the visualization of documents according to topics based on their content in an unsupervised way. That is, topics are not defined a priori. We performed this analysis using scikit-learn v1.0.2 \cite{scikit-learn}. More specifically, we selected the corpus of 15,359,185 tweets containing olive, coconut, and palm oil. Then, we obtained the 3,000 most representative words in this set and created a 15,359,185 $\times$ 3,000 matrix. Each term of the matrix contains the so-called \emph{term frequency-inverse document frequency} ($tf{-}idf$) that weights words according to how common they are in a document but penalizes those words that are very common across the whole set of documents. Lastly, we applied the LSA technique to reduce the number of dimensions to 2, which is the preferred choice for visualization purposes. Thus, it is, in general, not possible to associate a particular subject or discourse with them. See the Supplementary Information for further details.

\subsection*{Sentiment Analysis}
Sentiment Analysis is a supervised machine-learning technique that relies on a pre-trained model to determine if the expressed opinion in a document is positive, negative, or neutral. In particular, we employed the \emph{Twitter-xlm-roberta-base-sentiment} model. This model was pre-trained on a corpus of almost 200 million tweets in 30 languages and fine-tuned for sentiment analysis \cite{twittermodel}. For each tweet, the model assigns a score to the labels negative, neutral, and positive. For simplicity, we chose the label with the highest score to classify its sentiment. See the Supplementary Information for further details.

\subsection*{Virality characterization}

To characterize virality, we focused on two important characteristics of information cascades or avalanches. The first one is the interevent time, which measures the period between two tweets containing the same hashtag. The second one is the cascade size of a certain topic, which measures its popularity.

\noindent
\textbf{Interevent time (IET): } the IET associated with a certain hashtag is the distribution of the time gap between each pair of tweets containing the said hashtag. We fit these data to a power-law with a probability density function:
\begin{equation}
    \label{eq:pl}
    p(\tau) \propto \tau^{-\alpha},
\end{equation}
\noindent
where $\alpha$ is known as the scaling parameter or scaling exponent, which determines the characteristics of these heavy-tail distributions \cite{Clauset_2009}.

\noindent
\textbf{Cascade size (CS): } in our context, an information cascade can be broadly defined as the series of tweets with a given hashtag posted after an initial tweet containing it. We define the CS distribution of a certain hashtag as the distribution of the number of tweets posted every day using that hashtag. We fit these distributions using the same heavy-tailed distribution defined in eq.\eqref{eq:pl}.

\noindent
We fit the IET and CS distributions using the Powerlaw v1.5 package \cite{pwl} (see the Supplementary Information for further details).

\noindent
\textbf{Virality regions: } a power-law has a well-defined mean only if $\alpha > 2$, and it has well-defined variance only if $\alpha > 3$ \cite{Newman2005Sep}. This distinctive characteristic of these distributions allows us to group hashtags according to the specific values of their scaling exponent for the CS ($\alpha_\text{CS}$) and IET ($\alpha_\text{IET}$ distributions. In Fig.~\ref{fig:viral}, we observe that our hashtags may fall into 6 different regions:

\begin{itemize}
\item Region I ($2 \leq \alpha_\text{CS} < 3$ \& $\alpha_\text{IET} < 2$):  virally big, unpredictably fast regime. The average CS is finite, while the average and higher moments of the IET distribution diverge in the infinite size limit. A large variance of the CS signals that, even though the average is well-defined, there might be exceptionally large viral events. The large average and variance of the IET distribution indicate that these events may be very fast but rare.

\item Region II ($2 \leq \alpha_\text{CS} < 3$ \& $2 \leq \alpha_\text{IET} < 3$): viral regime. This is the typical signature of viral events. Both averages are finite, but the divergence of the variance (in the infinite size limit) indicates that they may be bursty and large.

\item Region III ($2 \leq \alpha_\text{CS} < 3$ \& $\alpha_\text{IET} \geq 3$): virally big, unvirally slow regime. The variance of the CS is large, indicating that there might be exceptionally large events, but they occur at predictable rates.

\item Region IV ($\alpha_\text{CS} \geq 3$ \& $\alpha_\text{IET} < 2$): unvirally small, unpredictable fast regime. In contrast with Region I, events in this category have a well-defined size and do not enter into the viral regime.

\item Region V ($\alpha_\text{CS} \geq 3$ \& $2 \leq \alpha_\text{IET} < 3$): unvirally small, virally fast regime. Events in this regime have well-defined cascade sizes, but they are not regularly active and a large amount of time may elapse between successive cascades of the same hashtag.

\item Region VI ($\alpha_\text{CS} \geq 3$ \& $ \alpha_\text{IET} \geq 3$): unviral regime. Both the average and variance of the distributions are finite, and thus, hashtags that lie in this regime can be labeled as non-viral.

\end{itemize}

\noindent
Note that we did not find any hashtag with $\alpha_\text{CS} < 2$ and, thus, we have removed those regions from the discussion to facilitate it.

\section*{Data \& Code Availability}

Twitter's Developer Agreement Policy does not allow users with Academic Research access to deposit the data in any public repository, but it allows them to share the ID of the tweets and the users analyzed without limits. These, together with the code necessary to reproduce the results, will be archived at zenodo and assigned a DOI upon acceptance.

\section*{Acknowledgements}
E.C., A.A., H.F.A., and Y.M. acknowledge the financial support of Soremartec S.A. and Soremartec Italia, Ferrero Group. Y.M. acknowledges partial support from the Government of Aragon and FEDER funds, Spain through grant E36-20R (FENOL), and by MCIN/AEI and FEDER funds (grant PID2020-115800GB-I00). A.A. acknowledges support from Grant RYC2021‐033226‐I funded by MCIN/AEI/10.13039/501100011033 and the European Union NextGenerationEU/PRTR. We also acknowledge support from Banco Santander (Santander-UZ 2020/0274). The funders had no role in the study design, data collection, analysis, decision to publish, or preparation of the manuscript.

\section*{Author contributions statement}

A.A., H.F.A., and Y.M. conceptualized the research. A.A. collected the data. E.C. implemented and carried out all the analyses. E.C., A.A., and H.F.A. wrote the initial draft. A.A., H.F.A., and Y.M. supervised the project. E.C., A.A., H.F.A., E. M, and Y.M. discussed and interpreted the results. All authors reviewed the manuscript. Y.M. acquired the funding necessary for the project.

\section*{Competing interests} 

The authors declare no competing interests.

\clearpage

\clearpage

\section*{Supplementary Material: Understanding the Vegetable Oil Debate and Its Implications for Sustainability through Social Media}

\section{Anatomy of the vegetable oils presence in Twitter}
\begin{figure}[h]
    \centering
    \includegraphics[width=0.6\textwidth]{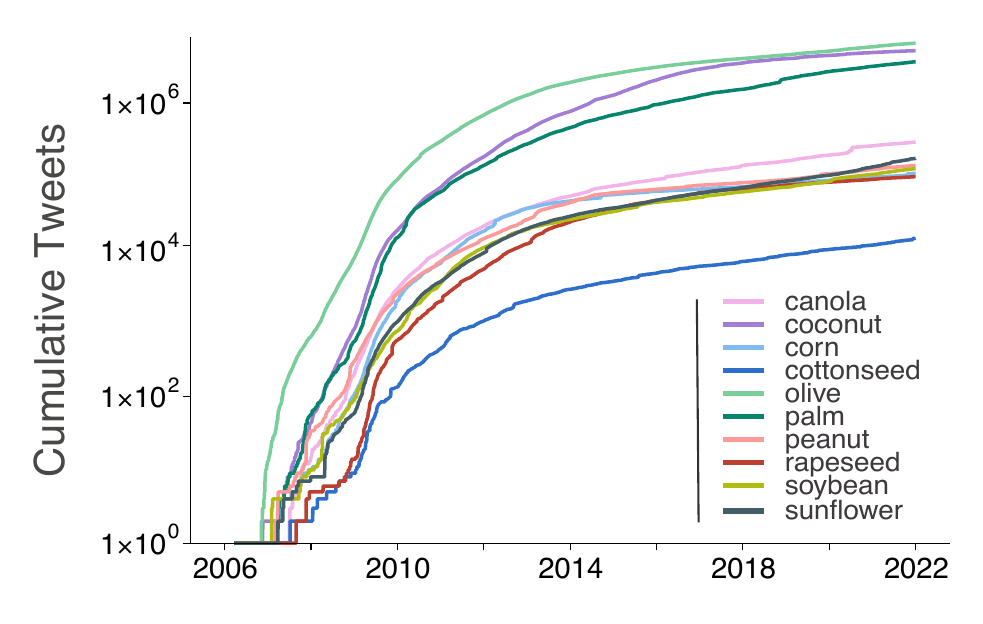}
    \caption{Cumulative number of tweets containing the bigram \emph{``type oil''}, where type is any of the ones present in the legend.}
    \label{fig:cumulative_log}
\end{figure}

Fig. \ref{fig:cumulative_log} shows the cumulative number of tweets containing the bigram ``\textit{type} oil'' where \textit{type} stands for each of the oils considered. Tab.~\ref{tab:cumulative} summarizes the total number of tweets obtained for each oil. We can clearly distinguish at least two groups. Olive, coconut, and palm oils are the major oils in terms of public debate on Twitter, with over 3,000,000 tweets for each oil in the last 15 years. In contrast, the number of tweets for the other oils is one order of magnitude smaller. The least discussed oil is cottonseed oil, even though the world supply in 2022 was 4.97 million metric tons, almost 50\% larger than coconut (3.59) or olive (3.29) oil \cite{oilSeeds}.

\begin{table}[htbp]
\centering
\begin{tabular}{@{}ll@{}}
\toprule
Bigram         & Tweets  \\ \midrule
olive oil      & 6,543,149 \\
coconut oil    & 5,165,202 \\
palm oil       & 3,650,834 \\
canola oil     & 293,361  \\
sunflower oil  & 176,152  \\
peanut oil     & 141,005  \\
soybean oil    & 127,654  \\
corn oil       & 109,895  \\
rapeseed oil   & 99,177   \\
cottonseed oil & 14,280   \\ \bottomrule
\end{tabular}
\caption{Cumulative number of tweets published between March 2006 and December 2021 with each bigram.}
\label{tab:cumulative}
\end{table}

\subsection{Granger causality test}

We test whether knowledge of the number of tweets containing common words is useful to predict the number of tweets from the three major oils. We apply the Granger causality test implemented in the library lmtest v0.9 in R v4.2.2, which consists of a Wald test. This test compares the unrestricted model (explaining the number of tweets from a certain oil, $o(t)$, using $o(t-lag)$, and the number of common words, $c(t-lag)$) with the restricted model (explaining $o(t)$ using only $o(t-lag)$). In all cases, we use the time series of monthly tweets and set as input lag 1 month. Tab. \ref{tab:granger} summarizes the results of the tests.

\begin{table}[htbp]
\centering
\begin{tabular}{@{}lll@{}}
\toprule
Oil     & F-test & p-value      \\ \midrule
olive   & 45.745 & \textless 0.001 \\
coconut & 2.8338 & 0.09398      \\
palm    & 43.502 & \textless 0.001 \\ \bottomrule
\end{tabular}
\caption{Results of the Granger causality test.}
\label{tab:granger}
\end{table}

\subsection{Palm oil in different languages}

To explore the debate on palm oil outside of the Anglosphere, we translated the bigram ``palm oil'' to over 100 languages using the library \emph{googletrans} v3.0 \cite{googletrans} with Python 3.8. After removing some duplicates (such as Spanish and Galician) we obtained 102 queries. In Tab. \ref{tab:language}, we show the complete list of queries together with the number of tweets containing the corresponding term from March 2006 to December 2021.
\begin{table}[h]
    \centering
\includegraphics[width=\textwidth]{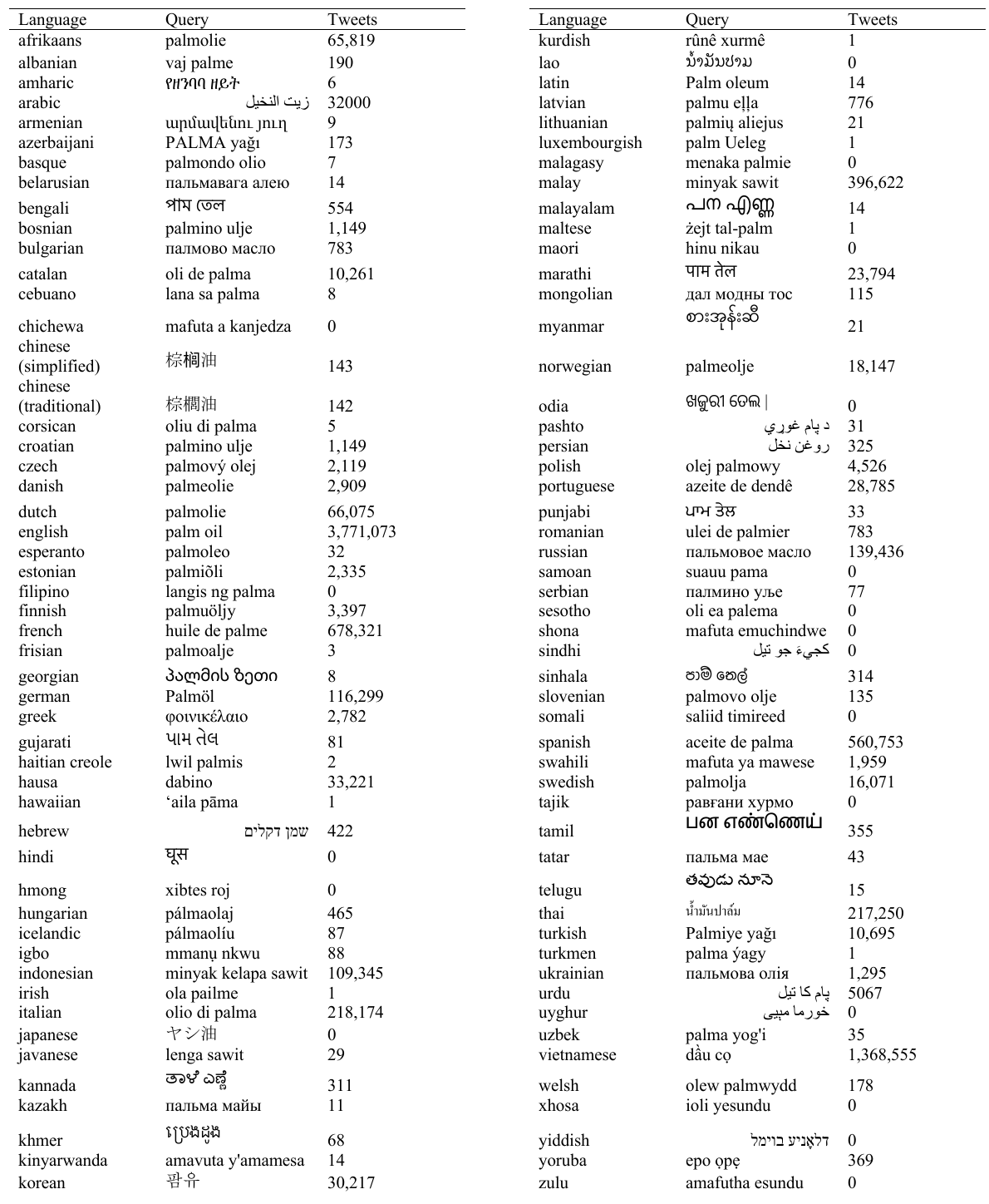}
    \caption{Query used to search for palm oil tweets in each language, together with the number of tweets obtained.}
    \label{tab:language}
\end{table}

\subsection{Common words}

\begin{figure}[h]
    \centering
    \includegraphics[width=0.6\textwidth]{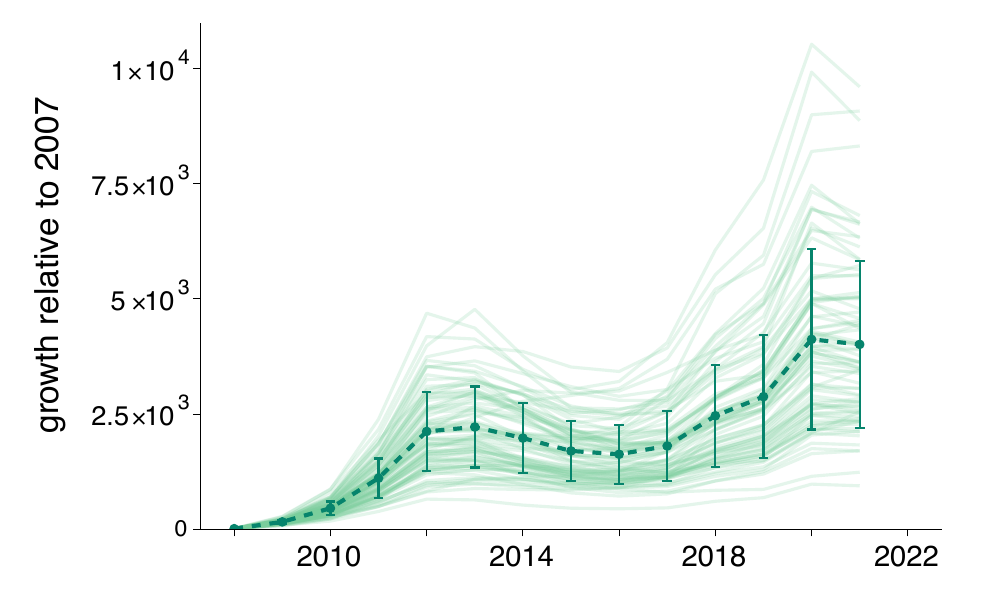}
    \caption{Relative growth of tweets containing common words, with 2007 as reference. Each line corresponds to one of the words shown in Tab. \ref{tab:words}, while the dashed line and whiskers indicate the average and standard deviation of the whole set.}
    \label{fig:words}
\end{figure}

To estimate the natural growth of tweets caused by the increase in the number of users using the platform, we obtained the number of tweets containing any of the 100 most common words in English. We rely on the list present in Wikipedia \cite{englishWiki}, which is based on an analysis of the Oxford English Corpus. Unfortunately, Twitter does not support queries that contain only a stop-word, and several words of this set are considered stop-words by Twitter. In particular, 19 words were considered stop-words by the API. Thus, the subset of common words contains information on the other 81. In Tab. \ref{tab:words}, we report the list of words used and the total volume of tweets published between March 2006 and December 2021. In Fig. \ref{fig:words}, we show the relative growth of tweets containing common words, together with their average and standard deviation.

\begin{table}[]
\begin{minipage}{0.24\textwidth}
\begin{tabular}{@{}ll@{}}
\toprule
Type    & Tweets (millions) \\ \midrule
a       & -                 \\
about   & 8,913             \\
after   & 3,620             \\
all     & 14,627            \\
also    & 2,205             \\
an      & -                 \\
and     & -                 \\
any     & 2,460             \\
as      & 10,666            \\
at      & -                 \\
back    & 5,554             \\
be      & 25,525            \\
because & 3,528             \\
but     & -                 \\
by      & -                 \\
can     & 13,810            \\
come    & 3,590             \\
could   & 2,747             \\
day     & 7,281             \\
do      & 15,910            \\
even    & 3,839             \\
first   & 3,909             \\
for     & 37,412            \\
from    & -                 \\
get     & 10,269            \\ \bottomrule
\end{tabular}

\end{minipage} \hfill
\begin{minipage}{0.24\textwidth}
\begin{tabular}{@{}ll@{}}
\toprule
Type    & Tweets (millions) \\ \midrule
give    & 2,424             \\
go      & 7,475             \\
good    & 7,820             \\
have    & 15,018            \\
he      & 10,635            \\
her     & 5,237             \\
him     & 3,997             \\
his     & 6,294             \\
how     & 8,647             \\
I       & 70,984            \\
if      & -                 \\
in      & -                 \\
into    & 2,670             \\
it      & -                 \\
its     & -                 \\
just    & 14,666            \\
know    & 7,133             \\
like    & 13,433            \\
look    & 3,288             \\
make    & 5,011             \\
me      & -                 \\
most    & 2,610             \\
my      & -                 \\
new     & 8,614             \\
no      & 32,162            \\ \bottomrule
\end{tabular}

\end{minipage} \hfill
\begin{minipage}{0.24\textwidth}
\begin{tabular}{@{}ll@{}}
\toprule
Type    & Tweets (millions) \\ \midrule
not     & 14,097            \\
now     & 9,636             \\
of      & 40,685            \\
on      & 30,561            \\
one     & 10,496            \\
only    & 4,895             \\
or      & -                 \\
other   & 2,575             \\
our     & 6,507             \\
out     & 10,526            \\
over    & 4,050             \\
people  & 7,782             \\
say     & 3,694             \\
see     & 6,002             \\
she     & 5,051             \\
so      & 19,866            \\
some    & 5,026             \\
take    & 3,434             \\
than    & 3,645             \\
that    & 25,455            \\
the     & -                 \\
their   & 4,871             \\
them    & 4,937             \\
then    & 3,497             \\
there   & 5,911             \\ \bottomrule
\end{tabular}

\end{minipage} \hfill
\begin{minipage}{0.24\textwidth}
\begin{tabular}{@{}ll@{}}
\toprule
Type    & Tweets (millions) \\ \midrule
these   & 3,231             \\
they    & 10,101            \\
think   & 4,568             \\
this    & -                 \\
time    & 7,768             \\
to      & -                 \\
two     & 2,000             \\
up      & 11,089            \\
us      & 6,945             \\
use     & 1,600             \\
want    & 5,578             \\
way     & 3,527             \\
we      & -                 \\
well    & 3,167             \\
what    & 12,129            \\
when    & 9,604             \\
which   & 1,711             \\
who     & 8,458             \\
will    & 9,779             \\
with    & 20,237            \\
work    & 3,193             \\
would   & 4,554             \\
year    & 3,415             \\
you     & -                 \\
your    & 15,000            \\ \bottomrule
\end{tabular}
\end{minipage}
\caption{List of common words used to estimate Twitter's growth, together with the total number of tweets posted between March 2006 and December 2021. Empty values mark stop-words that cannot be queried alone.}
\label{tab:words}
\end{table}

\subsection{Relative growth}

\begin{figure}[h]
    \centering
    \includegraphics[width=\textwidth]{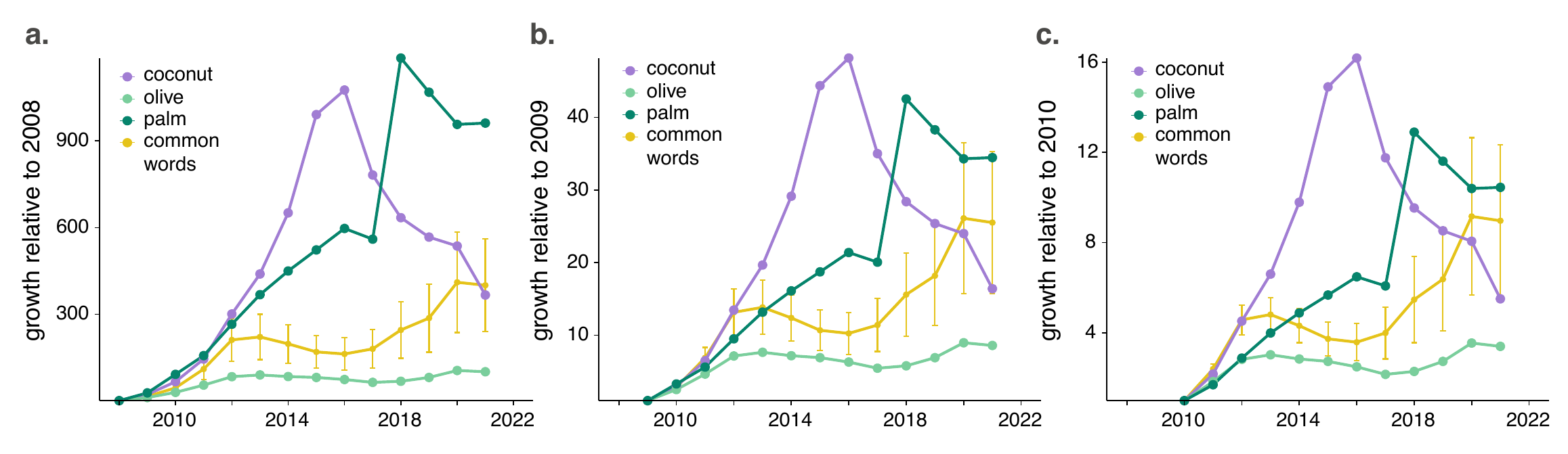}
    \caption{Relative growth of tweets using as reference years 2008 (a), 2009 (b) and 2010 (c).}
    \label{fig:growth}
\end{figure}

In Fig. \ref{fig:growth}, we show the relative growth in the number of tweets using different years as reference. We observe that the overall shape and conclusions do not depend on this choice.

\section{Natural Language Processing}

We characterize the content of tweets using two natural language processing (NLP) techniques. The first one, latent semantic analysis (LSA), is an unsupervised machine learning technique that projects documents (embedding them in a low-dimensional space) according to their topic. The second one, sentiment analysis, is a supervised machine-learning technique that relies on a pre-trained model to determine if the expressed opinion in a document is positive, negative or neutral.

\subsection{Latent Semantic Analysis}

We simplify the representation of the set of tweets using a bag-of-words model\cite{Bunt1999-BUNCM-4}. With this model, each document (each tweet) is represented by the multiset of its words, disregarding grammar and word order. With these multisets, we can build the document-term matrix
\begin{equation}
M_{i,j} = w_{i,j},
\end{equation}
\noindent
where $w_{i,j}$ is the frequency of word $j$ in document $i$. These weights can be further refined to properly reflect how important a word is to a document in a corpus. To do so, we use the common technique known as \emph{term frequency-inverse document frequency} ($tf{-}idf$). The term frequency is defined as:
\begin{equation}
    tf_{i,j} = \frac{n_{i,j}}{|d_i|},
\end{equation}
\noindent
where $n_{i,j}$ is the number of occurrences of the word $j$ in the document $i$, and $|d_i|$ is the length of the document (measured in words). Conversely, there are multiple definitions of the inverse document frequency. In this work, we adopt the implementation of scikit-learn v1.0.2 \cite{scikit-learn} in which the $idf$ of document $j$ is given by:
\begin{equation}
    idf_{j} = \log_{10} \frac{|D|}{1 + |\{d\in D: j\in d\}|},
\end{equation}
where $|D|$ is the size of the complete set of documents, and $|\{d \in D: j\in d\}|$ is the number of documents containing the word $j$. This term penalizes words that are very common in the corpus. 

With these terms, the $tf{-}idf$ matrix can be computed as:
\begin{equation}
    tf{-}idf_{i,j} = tf_{i,j} \times idf_j ,
\end{equation}
\noindent
so that words that are very common in a document but hardly found in the corpus score much higher. In contrast, words that are very common across texts will have a large $idf$ and, thus, their $tf{-}idf$ will tend to 0. Therefore, the weights in the $tf{-}idf$ matrix capture the importance of each word in characterizing such a document.

We can then apply the LSA technique to reduce this high-dimensional matrix while preserving the most important characteristics of the documents. In particular, if we denote the $tf{-}idf$ matrix as $M'$, its singular value decomposition is:
\begin{equation}
    M' = U \Sigma V^T
\end{equation}
\noindent
where $U$ and $V$ are two orthogonal matrices and $\Sigma$ a diagonal matrix. We can then select the $d$ largest values of $\Sigma$ (known as singular values) to obtain a $d-$dimensional representation of the documents, making it easier to group them according to their similarity.

For the particular case of our dataset, the complete analysis can be summarized in:
\begin{enumerate}
    \item \emph{Text preprocessing:} for each of the 15,359,185 tweets, we converted the text into lowercase, and tokenized the words using the Natural Language Toolkit v.3.7 \cite{bird2009natural}. Then, we removed all stopwords. Lastly, we grouped together the inflected forms of each word using the WordNet lexical database \cite{wordnet}. This procedure, commonly known as lemmatization, allows us to analyze all inflected forms of a term as a single item. For example, the words \textit{talk} and \textit{talking} would both be represented by the same lemma, which is \textit{talk}. 

    \item \emph{tf-idf vectorization:} we used the method $TfidfVectorizer$ from scikit-learn. We set $min\_df = 10$ and $max\_df = 0.7$ to remove those words that are very uncommon or extremely common, respectively. We also set $max\_features=3,000$ to limit the vocabulary. This procedure yielded a 15,359,185 $\times$ 3,000 matrix.

    \item{Latent semantic analysis:} we used the method $TruncatedSVD$ from scikit-learn with $n\_components=2$ to reduce the dimensionality of the matrix to 2. This method returned the matrices $U$ and $V$, with dimensions $3,000\times 2$ and $2\times 15,359,185$, respectively.

    \item{Visualization:} the $V^T$ matrix gives the coordinates in the lower 2-dimensional space of all the tweets. These are represented in Fig. 2 of the main text.
\end{enumerate}

\subsection{Sentiment Analysis}

Domain-specific language models are more suited to handle social media than those trained on general-domain corpora \cite{barbieri2021xlmtwitter}. In the case of Twitter, elements such as usernames, hyperlinks, or hashtags have a particular role in the document. As such, we choose the \emph{Twitter-xlm-roberta-base-sentiment} model \cite{twittermodel}. This model, based on the more general XLM-RoBERTa \cite{conneau2019unsupervised}, was pre-trained on a corpus of almost 200 million tweets in 30 languages and fine-tuned for sentiment analysis. We then use the Transformers v4.25.1 library \cite{Transformers} to assign sentiment labels to our corpus of tweets with the model.

The procedure starts with the pre-processing of tweets. For this specific model, it is necessary to substitute any username mentioned with @ with ``@user'' and any hyperlink with just ``http''. The model recognizes these particular structures and treats them accordingly. Running the classification task on each tweet provides a set of values associated with each sentiment:
\begin{equation}
    \text{scores}(t) = \{s_t(i)\}_{i=0,1,2} ,
\end{equation}
\noindent
where $s_t(i)$ is the score of label $i$ (0 stands for negative, 1 for neutral, and 2 for positive) for tweet $t$. We simplify this description and assign the label with the highest score to each tweet so that the label $l$ of tweet $t$ is
\begin{equation}
    l(t) = \underset{i \in \{0,1,2\}}{\mathrm{argmax}}\;\;\;scores(t) .
\end{equation}

\begin{figure}[h]
    \centering
    \includegraphics[width=\textwidth]{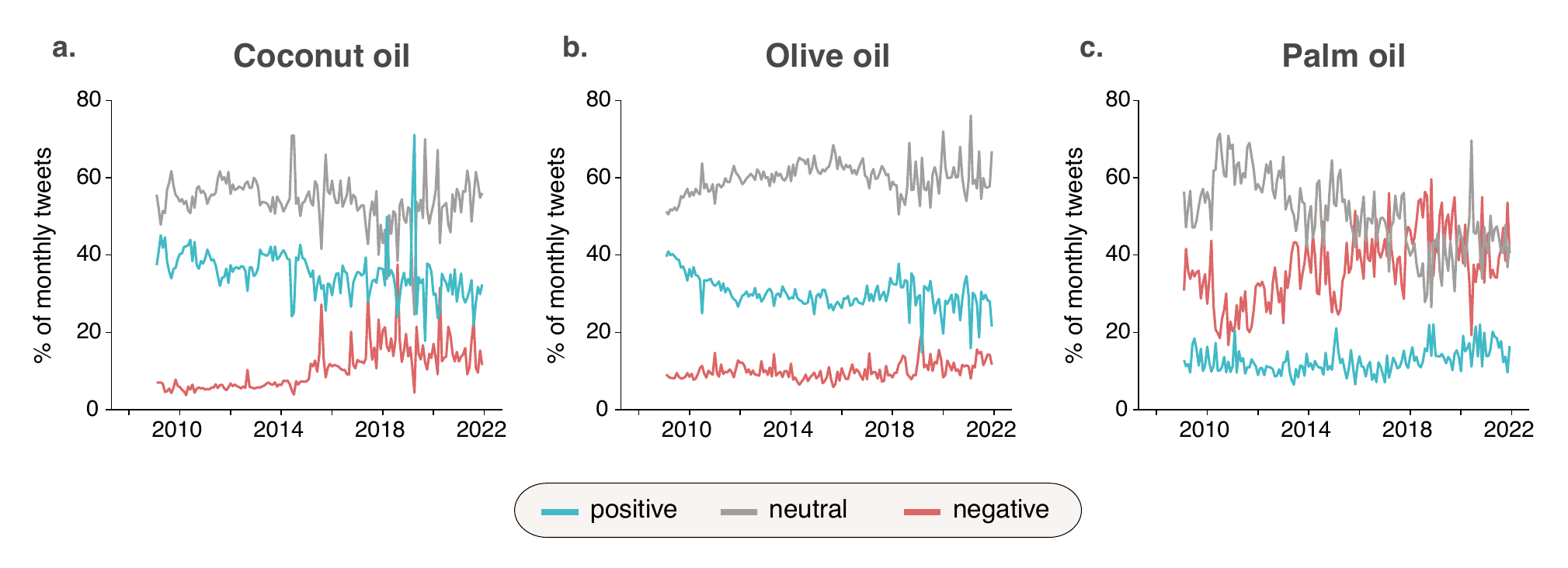}
    \caption{Evolution of the percentage of monthly tweets associated with each sentiment.}
    \label{fig:sentiment_distribution}
\end{figure}

Our results indicate that the sentiment distribution is fairly constant for the coconut and olive oil datasets. In contrast, the proportion of negative sentiments associated to palm oil constantly increases since 2010 at the expense of neutral tweets, Fig. \ref{fig:sentiment_distribution}. After a crossover in 2018, the fraction of tweets associated with negative sentiments is usually the dominant one.

\section{Virality characterization}

We fit the interevent time (IET) and cascade size (CS) distributions using the Powerlaw v1.5 package \cite{pwl}. Heavy-tailed distributions are characterized by their tail and, thus, it is possible to discard the smaller values of the data if they do not follow the distribution. To do so, this package implements the following procedure:
\begin{enumerate}
    \item Create a power-law fit starting from each unique value in the dataset ($\tau_\text{min}$).
    \item Select the one that results in the minimal Kolmogorov-Smirnov distance, $D$, defined as
    \begin{equation}
        D = \max_{\tau \geq \tau_{min}} |S(\tau)-P(\tau)|
    \end{equation}
    where $S(\tau)$ is the CDF for the empirical data, while $P(\tau)$ is the CDF for the heavy-tailed distribution we are considering as a hypothesis.
    \item Fit the data starting from a said value of $\tau$.
\end{enumerate}

\begin{figure}[h]
    \centering
    \includegraphics[width=\textwidth]{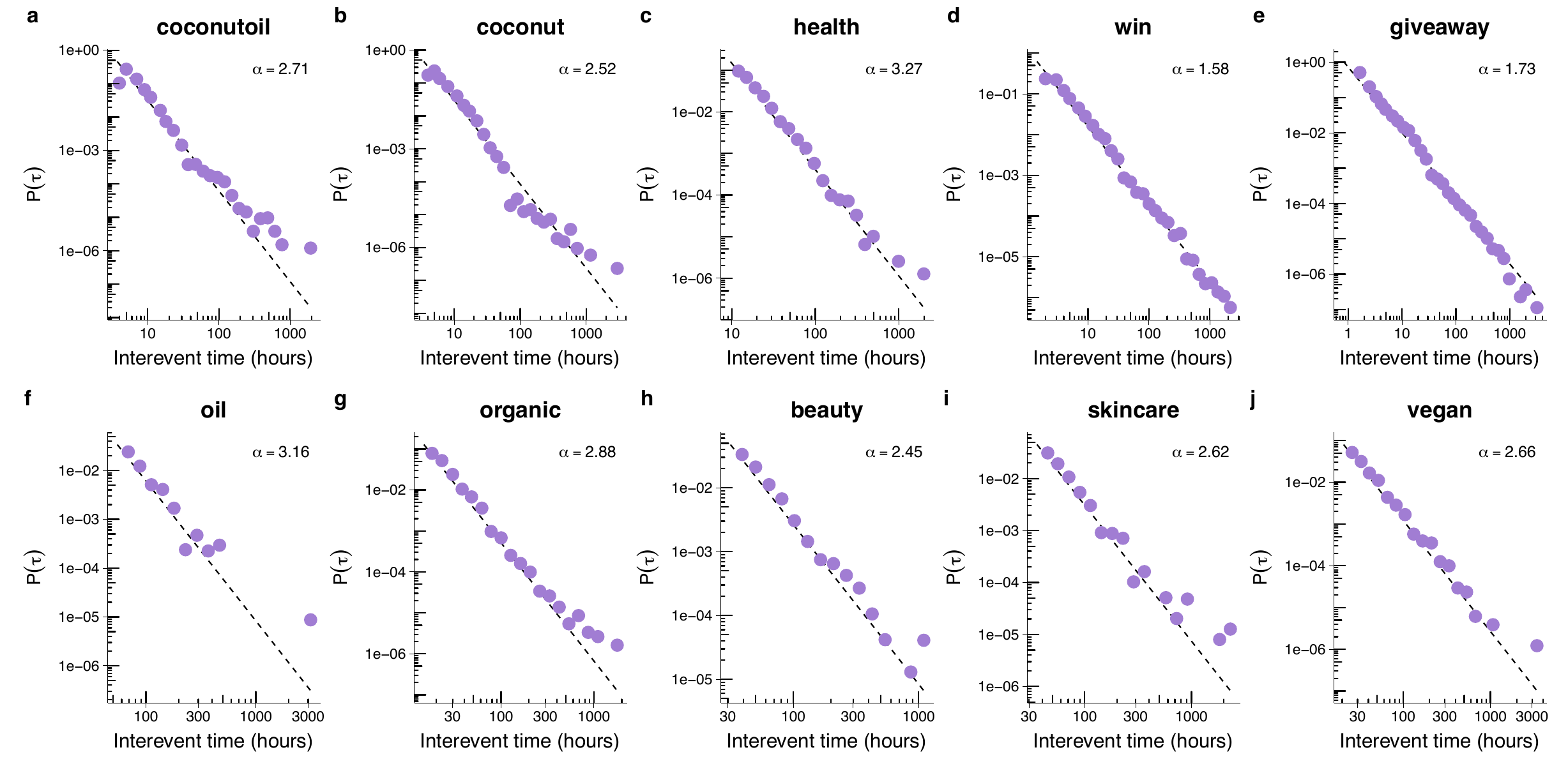}
    \caption{Fitting the IET distribution to a power-law distribution for the coconut oil dataset.}
    \label{fig:fit_IET_coconut}
\end{figure}

\begin{figure}[h]
    \centering
    \includegraphics[width=\textwidth]{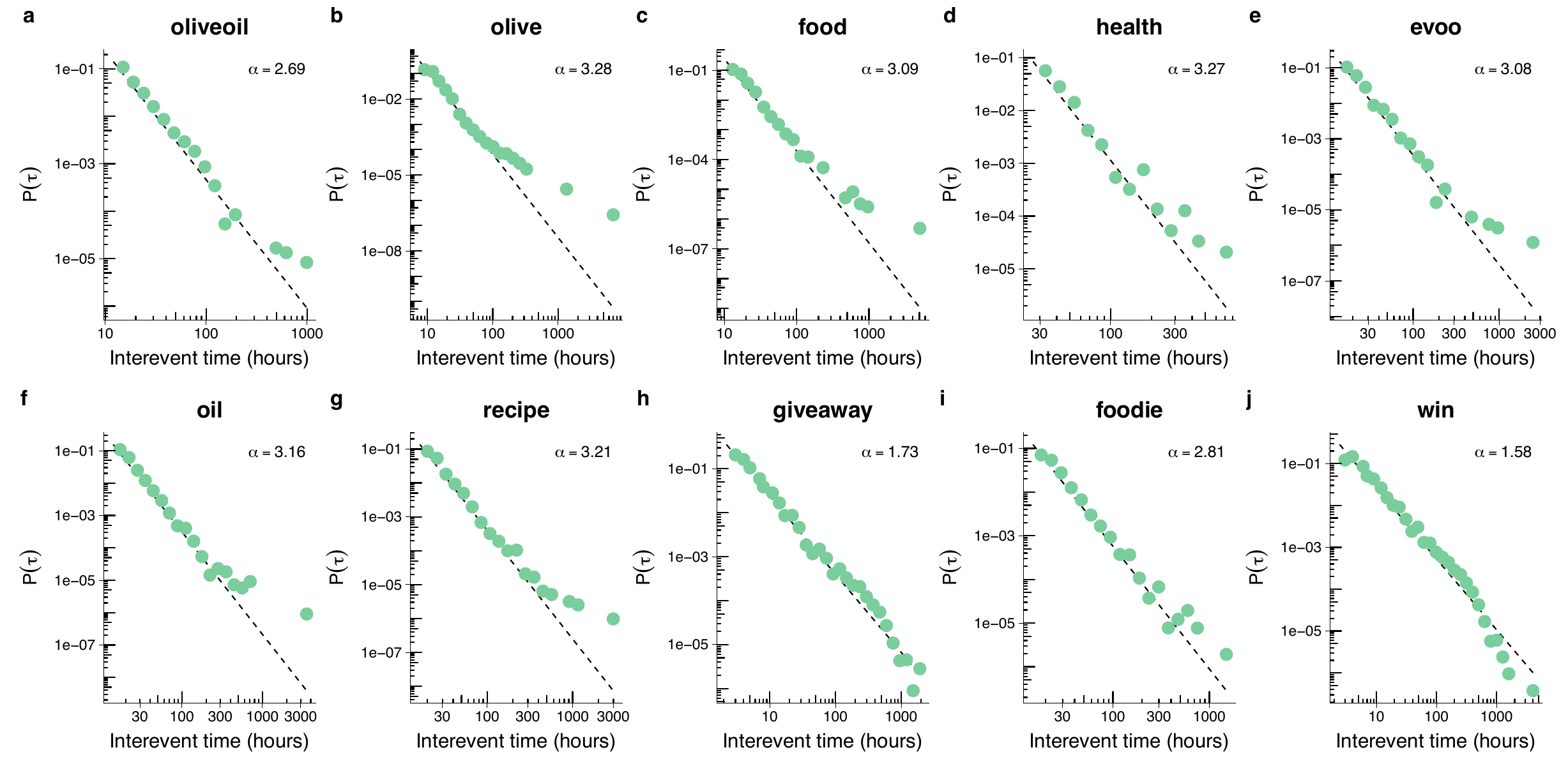}
    \caption{Fitting the IET distribution to a power-law distribution for the olive oil dataset.}
    \label{fig:fit_IET_olive}
\end{figure}

\begin{figure}[h]
    \centering
    \includegraphics[width=\textwidth]{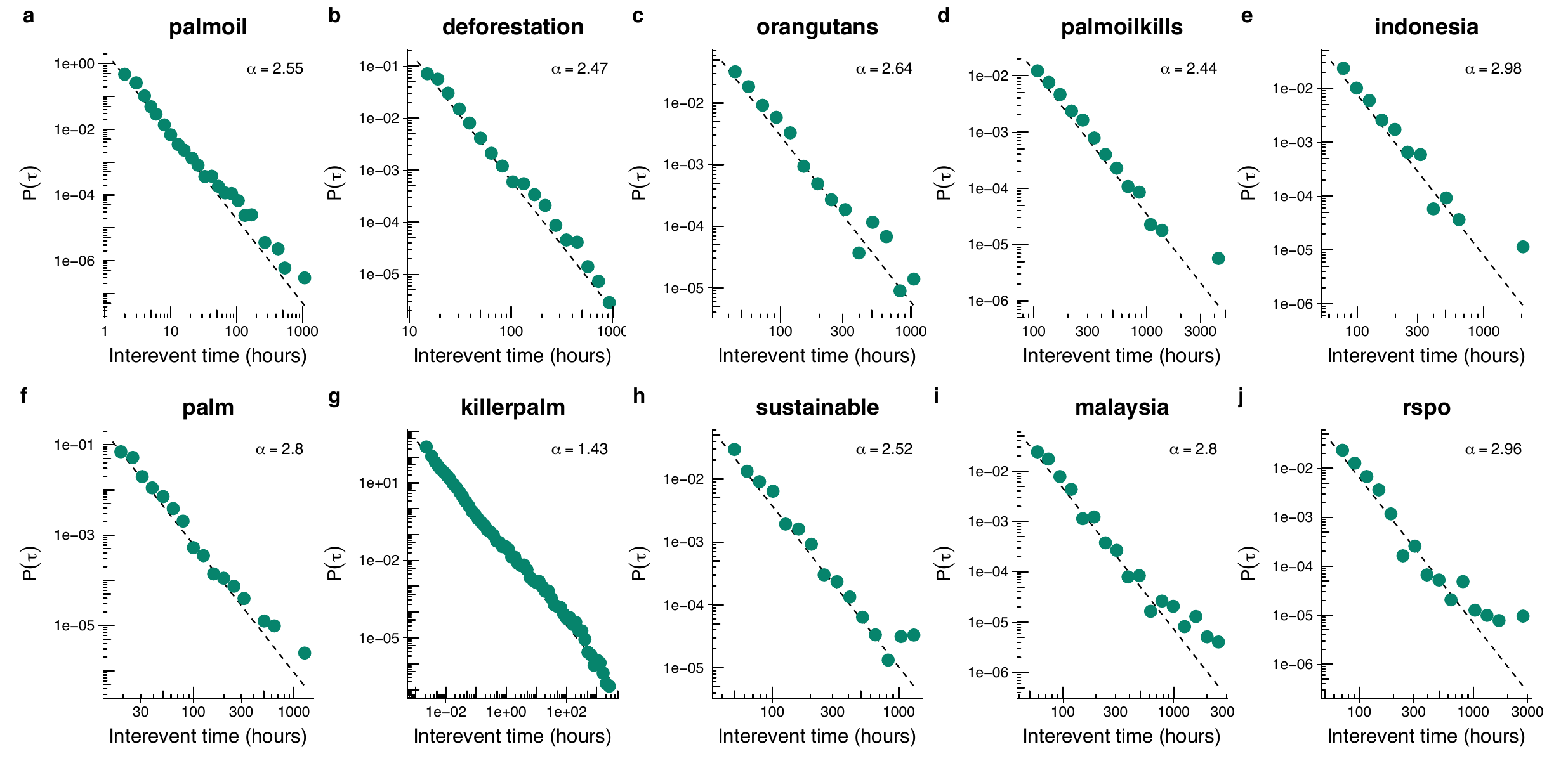}
    \caption{Fitting the IET distribution to a power-law distribution for the palm oil dataset.}
    \label{fig:fit_IET_palm}
\end{figure}

\begin{figure}[h]
    \centering
    \includegraphics[width=\textwidth]{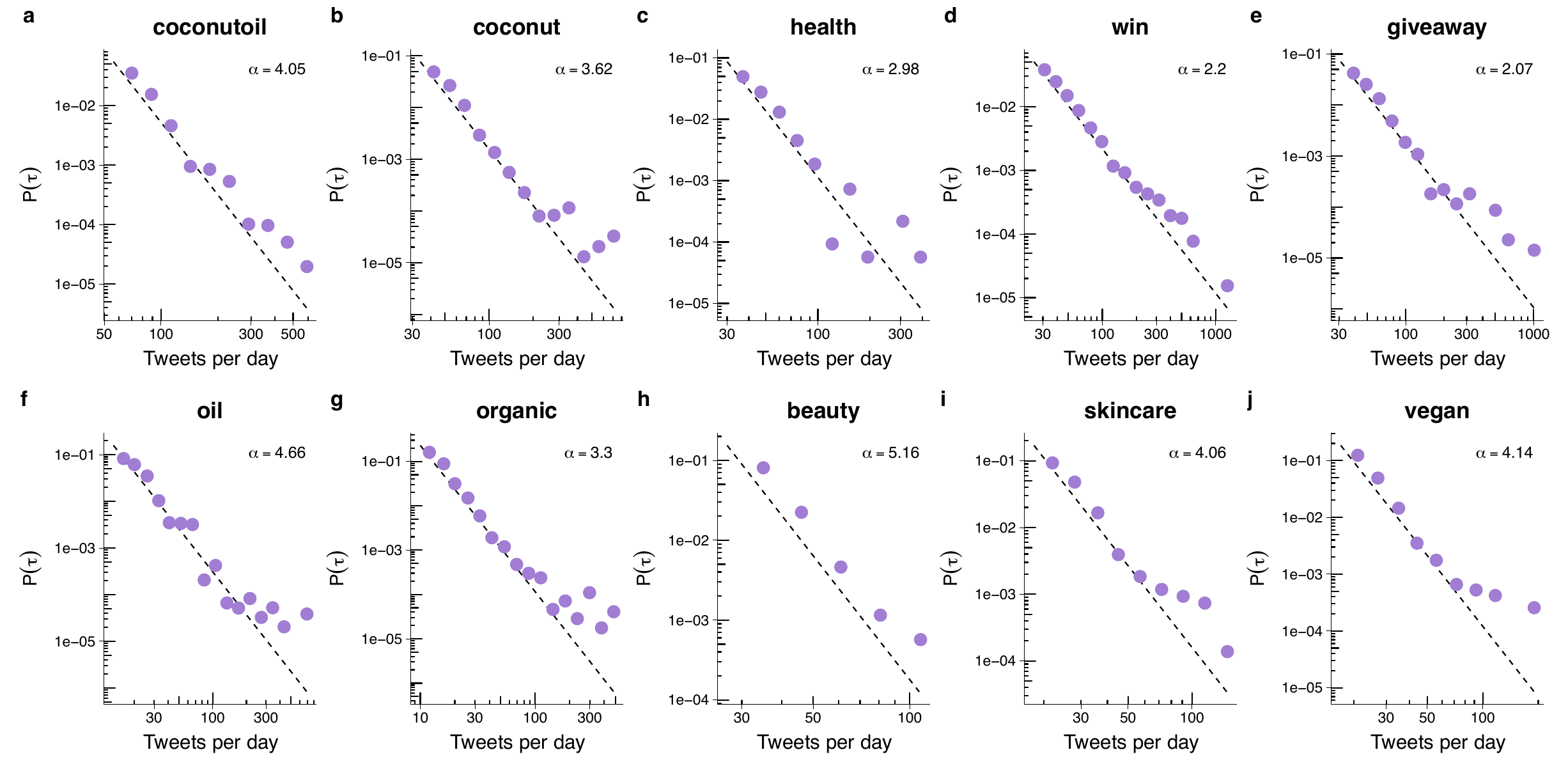}
    \caption{Fitting the CS distribution to a power-law distribution for the coconut oil dataset.}
    \label{fig:fit_CS_coconut}
\end{figure}

\begin{figure}[h]
    \centering
    \includegraphics[width=\textwidth]{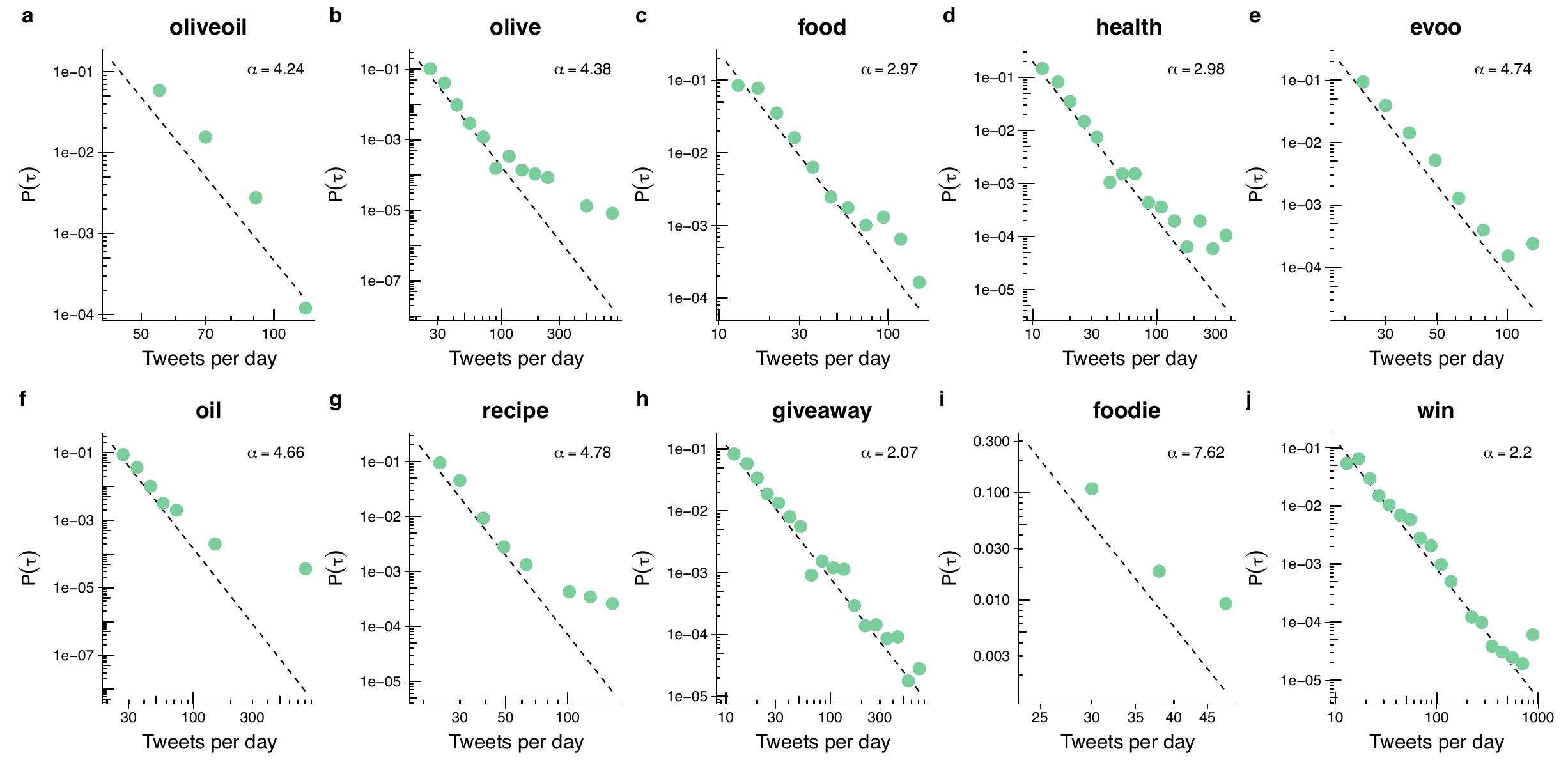}
    \caption{Fitting the CS distribution to a power-law distribution for the olive oil dataset.}
    \label{fig:fit_CS_olive}
\end{figure}

\begin{figure}[h]
    \centering
    \includegraphics[width=\textwidth]{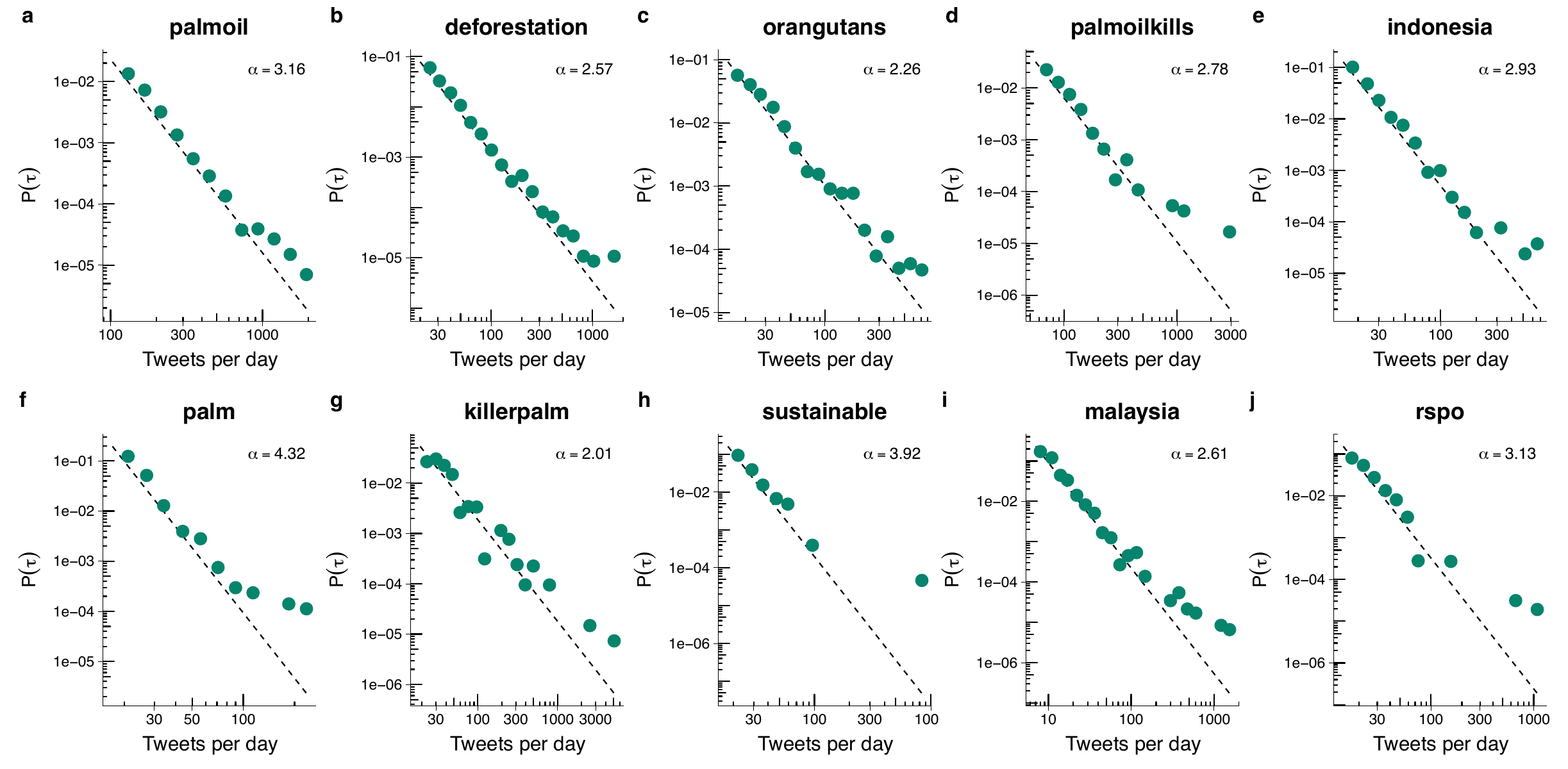}
    \caption{Fitting the CS distribution to a power-law distribution for the palm oil dataset.}
    \label{fig:fit_CS_palm}
\end{figure}

In figures \ref{fig:fit_IET_coconut}-\ref{fig:fit_IET_palm}, we show the fitted distributions to the IET distributions of each hashtag for the coconut, olive and palm oils, respectively. Similarly, in figures \ref{fig:fit_CS_coconut}-\ref{fig:fit_CS_palm}, we show the fitted distributions to the CS distributions. Tables \ref{tab:fit_IET_coconut}-\ref{tab:fit_IET_palm} show the results of the fitting procedure for the IET case and tables \ref{tab:fit_CS_coconut}-\ref{tab:fit_CS_palm} for the CS observable.

\begin{table}[]
\centering
\begin{tabular}{@{}lllll@{}}
\toprule
Hashtag    & $\tau_\text{min}$ & $\alpha$ \\ \midrule
coconutoil & 3.7               & 2.71     \\
coconut    &  3.5               & 2.52     \\
health     &  9.6               & 2.53     \\
win        &  1.5               & 1.87     \\
giveaway   &  0.8               & 1.81     \\
oil        &  54.9              & 2.87     \\
organic    &  14.7              & 2.88     \\
beauty     &  31.5              & 2.45     \\
skincare   &  35.8              & 2.62     \\
vegan      &  20.8              & 2.66     \\ \bottomrule
\end{tabular}
\caption{Fitting the IET distribution for hashtags related to coconut oil. }
\label{tab:fit_IET_coconut}
\end{table}

\begin{table}[]
\centering
\begin{tabular}{@{}lllll@{}}
\toprule
Hashtag  &  $\tau_\text{min}$ & $\alpha$ \\ \midrule
oliveoil &  12                & 2.69     \\
olive    &  7.7               & 3.28     \\
food     &  10.7              & 3.09     \\
health   &  26.8              & 3.27     \\
evoo     &  14                & 3.08     \\
oil      &  14.1              & 3.16     \\
recipe   &  16.4              & 3.21     \\
giveaway &  2.2               & 1.73     \\
foodie   &  14.8              & 2.81     \\
win      &  2.5               & 1.58     \\ \bottomrule
\end{tabular}
\caption{Fitting the IET distribution for hashtags related to olive oil.}
\label{tab:fit_IET_olive}
\end{table}

\begin{table}[]
\centering
\begin{tabular}{@{}lllll@{}}
\toprule
Hashtag       &  $\tau_\text{min}$ & $\alpha$ \\ \midrule
palmoil       &  1.3               & 2.55     \\
deforestation &  11.9              & 2.47     \\
orangutans    &  35.6              & 2.64     \\
palmoilkills  &  86.1              & 2.44     \\
indonesia     &  62.2              & 2.98     \\
palm          &  15.7              & 2.8      \\
killerpalm    &  0                 & 1.43     \\
sustainable   &  39.7              & 2.52     \\
malaysia      &  46.9              & 2.8      \\
rspo          &  56.3              & 2.96     \\ \bottomrule
\end{tabular}
\caption{Fitting the IET distribution for hashtags related to palm oil. }
\label{tab:fit_IET_palm}
\end{table}

\begin{table}[]
\centering
\begin{tabular}{@{}lllll@{}}
\toprule
Hashtag    & $\tau_\text{min}$ & $\alpha$ \\ \midrule
coconutoil &  56                & 4.05     \\
coconut    &  34                & 3.62     \\
health     &  30                & 3.60     \\
win        &  25                & 2.16     \\
giveaway   &  31                & 3.20     \\
oil        &  13                & 3.06     \\
organic    &  10                & 3.30     \\
beauty     &  27                & 5.16     \\
skincare   &  18                & 4.06     \\
vegan      &  17                & 4.14     \\ \bottomrule
\end{tabular}
\caption{Fitting the CS distribution for hashtags related to coconut oil. }
\label{tab:fit_CS_coconut}
\end{table}

\begin{table}[]
\centering
\begin{tabular}{@{}lllll@{}}
\toprule
Hashtag  &  $\tau_\text{min}$ & $\alpha$ \\ \midrule
oliveoil &  43                & 4.24     \\
olive    &  21                & 4.38     \\
food     &  11                & 2.97     \\
health   &  10                & 2.98     \\
evoo     &  19                & 4.74     \\
oil      &  22                & 4.66     \\
recipe   &  19                & 4.78     \\
giveaway &  10                & 2.07     \\
foodie   &  24                & 7.62     \\
win      &  11                & 2.20     \\ \bottomrule
\end{tabular}
\caption{Fitting the CS distribution for hashtags related to olive oil. }
\label{tab:fit_CS_olive}
\end{table}

\begin{table}[]
\centering
\begin{tabular}{@{}lllll@{}}
\toprule
Hashtag       &  $\tau_\text{min}$ & $\alpha$ \\ \midrule
palmoil       &  103               & 3.16     \\
deforestation &  20                & 2.57     \\
orangutans    &  14                & 2.26     \\
palmoilkills  &  56                & 2.78     \\
indonesia     &  15                & 2.93     \\
palm          &  17                & 4.32     \\
killerpalm    &  19                & 2.01     \\
sustainable   &  18                & 3.92     \\
malaysia      &  7                 & 2.61     \\
rspo          &  14                & 3.13     \\ \bottomrule
\end{tabular}
\caption{Fitting the CS distribution for hashtags related to palm oil. }
\label{tab:fit_CS_palm}
\end{table}

\end{document}